\documentclass[aps,preprint,showpacs]{revtex4}
\usepackage{amsfonts}
\usepackage{amsmath}
\usepackage{amssymb}
\usepackage{graphicx}
\usepackage{epsfig}

\begin{document}
\title{Transition states and the critical parameters of central potentials}

\author{Evgeny Z. Liverts}
\affiliation{Racah Institute of Physics, The Hebrew University, Jerusalem 91904,
Israel}

\author{Nir Barnea}
\affiliation{Racah Institute of Physics, The Hebrew University, Jerusalem 91904,
Israel}

\begin{abstract}

Transition states or quantum states of zero energy appear at the boundary
between the discrete part of the spectrum of negative energies and the 
continuum part of positive energy states.
As such, transition states can be regarded as a limiting case of a bound state with
vanishing binding energy, emerging for a particular set of critical potential parameters.
In this work we study the properties of these critical parameters for short range central potentials. 
To this end we develop two exact methods and also utilize the first and second order WKB approximations.
Using these methods we have calculated the critical parameters 
for several widely used central potentials.

The general analytic expressions for the asymptotic representations of the critical parameters
were derived for cases where either the orbital quantum number $l$ or the number $n$ of bound states approaches infinity.

The above mathematical models enable us to answer the following physical
(quantum mechanical) questions:
i) what is the number of bound states for a given central
potential and given orbital quantum number  $l$; 
ii) what is the maximum value of $l$ which can provide a bound
state for the given central potential;
iii) what is the order of energy levels for the given form of
the central potential.

It is revealed that the ordering of energy levels depends on the potential
singularity at the origin.
\end{abstract}


\maketitle
\section{Introduction}
\label{sec:introduction}
It is generally agreed that estimating the number of bound states of the
Schr\"{o}dinger 
equation is a problem of great practical importance. A substantial effort was
devoted for evaluating
the upper and lower limits on the number of bound states for a given central potential.
V. Bargmann \cite{BAR} and J. Schwinger \cite{SCHW} seem to be the first who tackled
this problem. Since then many authors have 
contributed to the study of this problem. Among them we
would like to refer to the works \cite{BER,SUB,FER1,FER2,DAI} and to emphasize
particularly the contribution of F. Calogero and F. Brau
\cite{CAL,BRU,CAB,CAB1,CAL2}.

The purpose of the current contribution is to present a systematic approach 
to the investigation of quantum states with zero energy.
These transition states appear between the discrete part of the spectrum of negative
energies and the continuum part of positive energies.
We study transition states by solving the proper Schr\"{o}dinger
equation for short range central potentials possessing specific sets of critical
parameters. 
To this end, we develop two exact methods for solving the zero energy Schr\"{o}dinger equation, and for obtaining the values of the associated critical parameters. 
We apply these methods to find the critical parameters of an improtant
class of central potentials including among others the Gaussian and the Yukawa interactions.
A few examples of potentials admitting the analytic solutions for the transition states
are exhibitted and the associated critical parameters are presented in analytic form.
We provide numerical results in the form of tables of the critical parameters.
These results enable one to obtain the exact number of bound states for these potentials without any additional computations. 

Analyzing our results we have observed the following universal properties of the
solutions of the Schr\"{o}dinger equation with central potentials:
(a) For a given orbital angular momentum quantum number $l$ 
the $n^{th}$ critical value of the universal parameter $\beta_{n,l}$ of any
potential behaves as $n^2$ for large $n$.  
(b) For a given number $n$ of bound states the critical parameter $\beta_{n,l}$ 
grows as $l^2$ with increasing $l$. Both results can be explained with the help of the WKB
approximation.
(c) The ordering of the energy levels with various $\{n,l\}$ depends on the potential
singularity at the origin.

The paper is organized as follows.
In  Section \ref{sec:shroedinger} we briefly discuss the Schr\"{o}dinger
equation and the properties of short range potentials.
The asymptotic behavior of the solution at the origin and at infinity is 
discussed in sections \ref{sec:origin},\ref{sec:asymptotic}, respectively.
In section \ref{sec:riccati} we introduce a phase-kind equation for the
transition states, and in section \ref{sec:wkb} we describe the WKB approximations 
for calculation of the critical
parameters. The analytic asymptotic expressions for the critical parameters
are presented in section \ref{sec:asymptotics}.
Numerical results and conclusions then follow in sections \ref{sec:results} and
\ref{sec:conclusions}.

\section{Short range potentials}
\label{sec:shroedinger}

Let us start this section with a quotation from section \S18
in the Landau \& Lifshitz textbook
on quantum mechanics \cite{LAN}: {\it ``If the field diminishes as $-1/r^s$  at infinity, with  $s>2$, then there are not levels of arbitrarily small negative energy.
The discrete spectrum terminates at a level with a non-zero absolute value, so that
the total number of levels is finite''}.

We notice, that the statement expounded in the first sentence of the above
quotation is doubtful. It contradicts our practical experience and also
section \S133 of the same book \cite{LAN}. The conclusion made in the second
sentence proves to be true. 

Therefore, we shall consider central potentials $V(r)$ satisfying the
corresponding boundary condition 
at infinity:
\begin{equation}
\label{1}
\lim_{r\rightarrow\infty}r^2 V(r)=0.
\end{equation}
Near the origin, the boundary condition for an \textit{attractive} potential
can be written in a similar form 
\begin{equation}
\label{2}
\lim_{r\rightarrow 0}r^2 V(r)=0,
\end{equation}
to avoid fall of a particle to the center \cite{LAN} (\S\S18,35).

We shall consider the class of potentials which can be presented in the form:
\begin{equation}
\label{3}
V(r,r_0)=-\dfrac{g}{r^s}f\left( \dfrac{r}{r_0}\right)~~~~~~~~~~~~~~~(g>0,~r_0>0),
\end{equation}
where $g$ is the coupling constant which determines the strength of the
interaction. 
Note that according to condition (\ref{2}) the power $s$ must
satisfy the inequality 
\begin{equation}
\label{3a}
q\equiv s+p<2,
\end{equation}
where $p$ corresponds to the leading term in series expansion of function $f(r)$ near the origin
\begin{equation}
  \label{4}
  f(r) \underset{r\rightarrow0}{\simeq} \dfrac{\overset{\sim}{g}}{r^p}.
\end{equation}
It should be realized that the definition (\ref{3}) describes a wide class of
potentials such as, e.g., square well, exponential, Hulthen, Gaussian, Yukawa,
Woods-Saxon and many others.

The radial part of the Schr\"{o}dinger equation for a  single particle moving
in a central potential field takes the form \cite{LAN}(\S32):
\begin{equation}
  \label{5}
  \dfrac{d^2\chi}{d r^2}+\left\lbrace\dfrac{2m}{\hbar^2}
  \left[ E-V(r) \right] -\dfrac{l(l+1)}{r^2} \right\rbrace \chi=0,
\end{equation}
where $m$ is the reduced mass, and $\chi\equiv\chi_l(r)$ is the reduced radial
part of the wave function for a
stationary state with angular momentum $l$ and energy $E$.

Changing the potential parameters the appearance of a new bound
state is accompanied with a new solution of
Eq.(\ref{5}) for $E=0$. 
As we limit our discussion to potentials of the form (\ref{3}), we rewrite Eq.(\ref{5}) for zero
energy in the form
\begin{equation}
\label{6}
\dfrac{d^2\chi(r)}{d r^2}=\left[-\dfrac{2m}{\hbar^2}
\dfrac{g}{r^s}f\left( \dfrac{r}{r_0}\right)
+\dfrac{l(l+1)}{r^2} \right]  \chi(r).
\end{equation}
The scale transformation $x=r/r_0$ leads to the equation:
\begin{equation}
\label{7}
\dfrac{d^2 \overset{\sim}{\chi}(x)}{d x^2}=\left[-\dfrac{2m}{\hbar^2}
\dfrac{g r_0^{2-s}}{x^s}f\left( x\right)
+\dfrac{l(l+1)}{x^2} \right] \overset{\sim}{\chi}(x).
\end{equation}
It is seen that the potentials $V(r,r_0)$ and $r_0^{2-s}V(r,1)$ have
equivalent solutions of Eq.(\ref{5}).
Thus, we are interested in solving the equation
\begin{equation}\label{8}
  \chi''(r)=U(r)\chi(r),
\end{equation}
where
\begin{equation}\label{9}
U(r)=-\beta v(r)+\frac{l(l+1)}{r^2},
\end{equation}
\begin{equation}\label{9a}
v(r)=f(r)r^{-s}.
\end{equation}
Now our choice of the form (\ref{3}) for central potentials becomes clear.
The solutions of Eq.(\ref{8}) for a given angular momentum quantum number $l$ 
depend effectively, (see, e.g.,\cite{PAT})
only on one parameter
\begin{equation}\label{10}
  \beta=\dfrac{2 m g}{\hbar^2} r_0^{2-s}.
\end{equation}
Eq.(\ref{8}) is a differential equation of a second order. In order to solve it both analytically and numerically one needs to know the behavior of its solution near the origin
and at infinity.

\section{The solution near the origin}
\label{sec:origin}
At first, let us consider the solution of Eq.(\ref{8}) near the origin.
Expanding the potential into a power series and keeping only the leading term 
we obtain 
\begin{equation}
\label{11}
\chi''(r)=\left[ -\dfrac{\lambda}{r^q}+\dfrac{l(l+1)}{r^2}\right] \chi(r),
\end{equation}
where
\begin{equation}
\label{12}
\lambda=\beta \overset{\sim}{g},
\end{equation}
and $q$ is defined by Eq.(\ref{3a}).
The particular solution of Eq.(\ref{11}), satisfying the boundary condition
\begin{equation}
\label{14}
\chi(0)=0 ,
\end{equation}
has the form
\begin{equation}
\label{15}
\chi(r)= A \sqrt{r}~J_{\frac{2l+1}{2-q}}\left(\dfrac{2\sqrt{\lambda}}{2-q}r^{1-q/2} \right),
\end{equation}
where $J_\nu(z)$ is the Bessel function of the first kind, and $A$ is an arbitrary constant.
Keeping the first two terms in the series expansion of the function (\ref{15}), one obtains
\begin{equation}\label{16}
  \chi(r)\underset{r\rightarrow0}{\simeq} A~r^{l+1}\left[(2-q)(2l+3-q)-\lambda~r^{2-q} \right].
\end{equation}
In general, for integer and half-integer $q$ the solution of Eq.(\ref{8}) satisfying the boundary condition (\ref{14}) can be presented by the following infinite series
\begin{equation}
\label{17}
\chi(r)=\overset{\sim}{A}~r^{l+1}\left[1+\sum_{i=1}^\infty r^i\left( b_i+r^{1-q}c_i\right)  \right].
\end{equation}
For integer $q$ all of the $b$-coefficients should be zero.
It follows from Eq.(\ref{16}) that
\begin{equation}
\label{18}
{\textstyle c}_1=-\dfrac{\lambda}{(2-q)(2l+3-q)}.
\end{equation}
Substituting the representation (\ref{17}) into Eq.(\ref{8}), and then equating the expansion coefficients of the same powers of $r$ for the left-hand (lhs) and right-hand sides (rhs)
of Eq.(\ref{8}), one can calculate any finite number of the subsequent coefficients $c_i$ and $b_i$ with $i\geq 1$.

\section{The asymptotic behavior of transition states}
\label{sec:asymptotic}

For an eigenfunction  $\Psi$ belonging to the discrete part of the spectrum, the integral
$\int\lvert\Psi\rvert^2 dV$,  taken over all space, is finite. This certainly
means that $\lvert\Psi\rvert^2$  decreases quite rapidly, becoming zero at
infinity. The motion of the system is limited to a finite range, and it is said
to be in a bound state. For wave functions belonging to the continuous part of
the spectrum, the integral $\int\lvert\Psi\rvert^2 dV$ diverges due to the fact
that $\lvert\Psi\rvert^2$ does not become zero at infinity
(or becomes zero insufficiently rapidly).

On the other hand \cite{LAN} (\S~18), the spectrum of negative eigenvalues of the energy is discrete, i.e. all states with $E<0$ in the field which vanishes at infinity are bound states. The positive eigenvalues $E>0$, on the other hand, form a continuous spectrum.

In other words, for the bound states the eigenvalues of the energy $E<0$, and the eigenfunctions must satisfy the boundary condition
\begin{equation}
\label{20}
\lim_{r\rightarrow \infty}\lvert\Psi\rvert^2=0.
\end{equation}
In contrary, for a free state that belongs to the continuous spectrum the energy eigenvalues $E>0$ and $\lvert\Psi\rvert^2$ does not become zero at infinity (or becomes zero insufficiently rapidly).

For the \textbf{transition states} with $E=0$ the asymptotic behavior ($r\rightarrow \infty$) of the eigenfunctions remains unclear.

The important step is to realize that the\textbf{ boundary condition} (\ref{20})
\textbf{must be valid for the transition states} ($E=0$) as well. Thus, for these states
$\lvert\Psi\rvert^2$ achieves zero at infinity. However, we note that it may tend to zero 
too slowly to ensure the convergence of the integral $\int\lvert\Psi\rvert^2 dV$.

In the following we shall rely on the boundary condition (\ref{20}) for the
transition states.

For $l>0$ the asymptotic boundary condition (\ref{1}) enables us to neglect the potential $V(r)=-\beta v(r)$ in Eqs.(\ref{8})-(\ref{9a}) at large enough $r$. The general solution of the resulting equation has a form
\begin{equation}
\label{21}
\chi_l(r)=C_1r^{l+1}+C_2 r^{-l},
\end{equation}
where $C_1$ and $C_2$ are arbitrary constants. As $\Psi \sim R(r)=\chi_l(r)/r$,
one should put $C_1=0$ in order to satisfy the asymptotic condition (\ref{20}). Thus, we get
\begin{equation}
\label{22}
\chi_l(r)\underset{r\rightarrow \infty}{\simeq}C_2 r^{-l}.
\end{equation}
Expressing the latter equation in the form $r^l \chi_l(r)\underset{r\rightarrow \infty}{\simeq} const$,
we obtain the following condition for the first derivative:
\begin{equation}
\label{23}
\lim_{r\rightarrow\infty}\dfrac{d}{dr}\left[r^l \chi_l(r) \right]\equiv
 \lim_{r\rightarrow\infty}\left[lr^{l-1}\chi_l(r)+r^l \chi'_l(r) \right]=0.
\end{equation}
It is clear that the asymptotic behavior of the solution $\chi_l (r)$ of Eq.(\ref{8}) depends on
the parameter $\beta$ of the effective potential (\ref{9}).
Thus, according to Eq. (\ref{23}),
the solution $\chi_l$ of Eq.(\ref{8}) fulfills the asymptotic condition
\begin{equation}
\label{24}
F_l(\beta_n)\equiv\lim_{r\rightarrow\infty}\left[ \dfrac{l}{r}\chi_l(r)+\chi'_l(r)\right]=0
\end{equation}
for the critical parameters $\beta_{n}$.
Here $n$ is a number of zeros of the function $F_l(\beta)$ for the given potential (\ref{9}).
Hence, by definition, if for a given $l$ the potential $V(r)$ is characterized by the parameters
meeting $\beta_{n+1}\geq\beta>\beta_n$ then the proper \textit{number of bound states} equals $n$.

\begin{figure}
\begin{center}
\caption{Function $F_0(\beta)=\lim_{r\rightarrow\infty}\chi_0'(r)$ for the Yukawa potential ($l=0$). Zeros of $F_0(\beta)$ present critical parameters $\beta_n$.} 
\epsfxsize=15cm\epsfbox{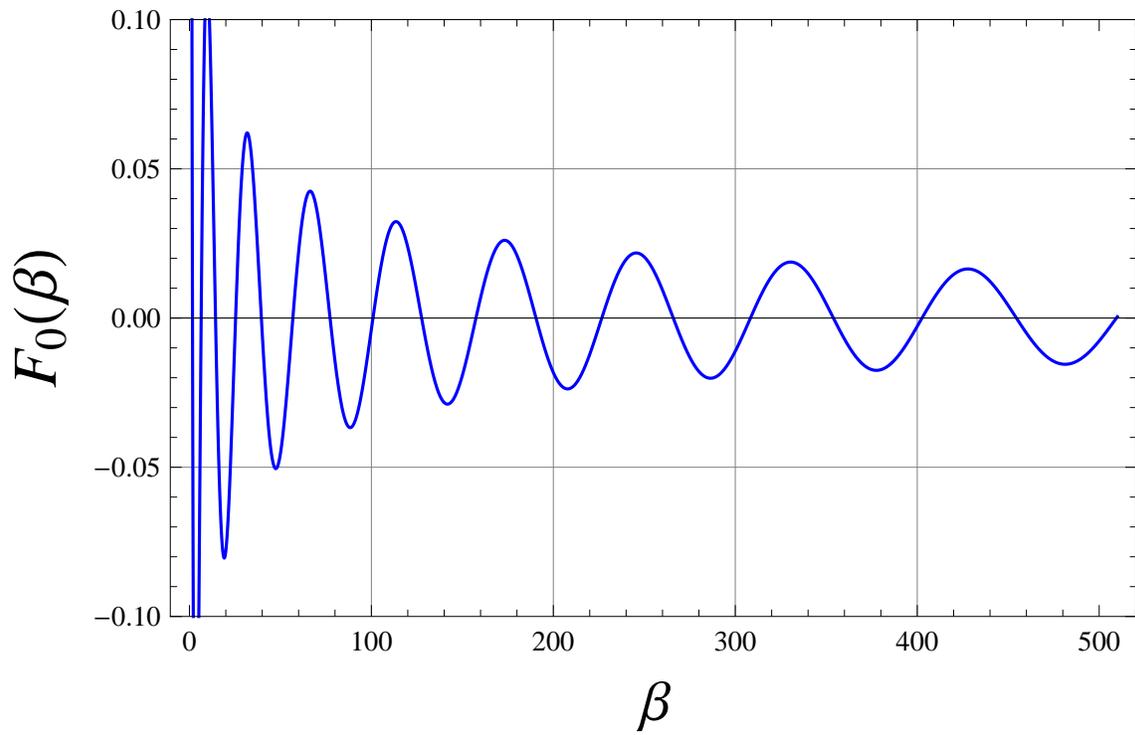} 
\label{F1}
\end{center}
\end{figure}

The asymptotic condition (\ref{24}) was derived assuming that
the orbital quantum number $l>0$. However, it is easy to show that Eq.(\ref{24}) preserves
its validity also for $l=0$. A typical graph of the function $F_l(\beta)$ is presented
in Fig. \ref{F1}.

The straightforward solution of the second order differential equation (\ref{8}) with
the boundary conditions (\ref{16}) and (\ref{24}) presents our \textbf{first method} for calculating
the critical parameters $\beta_n$ of a given attractive potential (\ref{3}) satisfying
the boundary conditions (\ref{1}) and (\ref{2}).
This method is especially effective and accurate for small values of $l$.
For a few potentials, such as the exponential, the Hulthen and the Woods-Saxon,
one can derive analytical expressions for the critical parameters
using this method (see Appendix). However, this is possible only for $S$-states ($l=0$),
when Eq.(\ref{8}) with the potentials mentioned above has a general analytical solution.
We are familiar with only one form of central potential which admits an analytical solution
of Eq.(\ref{8}) for $l\geq 0$. It is a cut-off potential (described in the Appendix) for which
the finite square well potential presents its particular case.

Let us add one important comment. We refer to Eq.(\ref{24}) as the asymptotic
behavior condition.
However, the solution (\ref{21}), and therefore  - condition (\ref{24}),
correspond to the assumption that the potential $V(r)$ is negligible in comparison with
the centrifugal term $l(l+1)/r^2$. Therefore, 
the condition (\ref{24}) is applicable at a distance $r$ when the condition
\begin{equation}
\label{25}
\lvert V(r)\lvert \ll \dfrac{l(l+1)}{r^2}
\end{equation}
is satisfied. 
\section{Equation of the phase kind}
\label{sec:riccati}
It was mentioned in the preceding section that the straightforward method for calculating
the critical parameters of central potentials loses its accuracy with increasing
angular momentum quantum number $l$. In this section we propose another method
for calculating these parameters. This method is
based on the logarithmic derivative $y(r)=\chi'(r)/\chi(r)$ of the reduced radial
wave function introduced earlier. The final equations are close to, but differ from the so called
phase equations presented in \cite{CAB,CAL2}.

Let us start with the trivial identity
\begin{equation}
\label{30}
\left( \dfrac{\chi'}{\chi}\right)'= \dfrac{\chi''}{\chi}-\left(\dfrac{\chi'}{\chi} \right)^2,
\end{equation}
and transform the radial Schr\"{o}dinger equation (\ref{8}) into a Riccati
type equation for the corresponding logarithmic derivative $y(r)\equiv y_l(r)$:
\begin{equation}
\label{31}
y'(r)+y^2(r)=U(r).
\end{equation}
The asymptotic behavior of the logarithmic derivative for $l>0$
\begin{equation}
\label{32}
y_l(r)\underset{r\rightarrow \infty}{\simeq}-\dfrac{l}{r}~~~~~~~~~~~~~~~~~~~~~~~~~~~(l>0)
\end{equation}
follows from the asymptotic representation (\ref{22}).

To deduce the asymptotic behavior of the logarithmic derivative for the transition $S$-states
($l=0$) let us start with the fact that for this case the rhs of Eq.(\ref{31}) equals $V(r)$.
Let us then consider central potentials with the asymptotic behavior
\begin{equation}
\label{33}
V(r)\underset{r\rightarrow \infty}{\simeq}-\dfrac{\beta}{r^\mu}~~~~~~~~~~~~(\beta>0)
\end{equation}
where $\mu>2$ according to the boundary condition (\ref{1}). The general solution of the proper
Schr\"{o}dinger equation (\ref{8}) has a form:
\begin{equation}
\label{34}
\chi(r)=\sqrt{r}\left[ C_2 J_\nu\left(2\nu\sqrt{\beta}r^{\frac{1}{2\nu}}\right)+
C_3 J_{-\nu}\left(2\nu\sqrt{\beta}r^{\frac{1}{2\nu}}\right)
 \right] ,
\end{equation}
with
\begin{equation}
\label{34a}
\nu\equiv\dfrac{1}{2-\mu}<0~~~~~~~~(\mu>2).
\end{equation}
It is seen that for $\mu>2$ the argument of the Bessel function goes to zero as $r\rightarrow \infty$.
Thus, using series expansion for the Bessel functions, it is easy to show that one should put
$C_2=0$ in order to satisfy the boundary condition (\ref{20}).
Taking then the logarithmic derivative for the resulting $\chi(r)$, and once more using a series expansion  for the Bessel functions, one obtains:
\begin{equation}
\label{34b}
y_0(r)\underset{r\rightarrow \infty}{\simeq}\dfrac{\beta}{\mu-1}r^{1-\mu}.
\end{equation}
It is clear that the asymptotic solution (\ref{34b}) of the Riccati equation (\ref{31}) for $l=0$
satisfies the following inequality:
\begin{equation}
\label{34c}
y_0^2(r)\underset{r\rightarrow \infty}{\ll} \lvert y_0'(r)\lvert.
\end{equation}
The asymptotic solution of the Schr\"{o}dinger equation (\ref{8}) with the exponential
potential is presented in the Appendix (see Eq.(\ref{A12a})) for $l=0$.
The corresponding logarithmic derivative 
\begin{equation}
\label{34d}
y_0^{exp}(r)\underset{r\rightarrow \infty}{\simeq}\beta e^{-r}
\end{equation}
satisfies the inequality (\ref{34c}).
It is easy to show that the asymptotic representation ($r\rightarrow \infty$) of the Hulthen and the Woods-Saxon potentials reduces to the exponential forms $-\beta e^{-r}$ and
$-\beta x_0^{-1}e^{-r}$, respectively.
Hence, the asymptotic solution of the corresponding Eq.(\ref{31}) can be presented
by the rhs of Eq.(\ref{34d}) for the Hulthen potential, and by $\beta x_0^{-1} e^{-r}$
for the Woods-Saxon potential. The latter logarithmic derivatives certainly obey the inequality (\ref{34c}) as well.

It is reasonable to suggest that inequality (\ref{34c}) is valid for all of
the short range potentials (may be excluding only the cut-off potentials).
In this case, one can neglect the square of the logarithmic derivative in the lhs of Eq.(\ref{31}). The solution of the latter equation with $l=0$ can be obtained then in the explicit form
\begin{equation}
\label{35}
y_0(r)\underset{r\rightarrow \infty}{\simeq}\beta\int_r^\infty v(r)dr.
\end{equation}
For potentials with the asymptotic behavior (\ref{33}), formula (\ref{35})
gives the asymptotic representation (\ref{34b}). For the exponential potential, the rhs of Eq.(\ref{35}) leads to (\ref{34d}).

For the Yukawa and Gaussian potentials Eq.(\ref{35}) yields:
\begin{equation}
\label{36}
y_0^{Yuk}(r)\underset{r\rightarrow \infty}{\simeq}\beta\dfrac{e^{-r}}{r},~~~~~~~~~~~~~
y_0^{Gau}(r)\underset{r\rightarrow \infty}{\simeq}\dfrac{\beta}{2}\dfrac{e^{-r^2}}{r}.
\end{equation}
For deriving the latter expressions we used the leading terms of the asymptotic expansions
of the incomplete gamma function $\Gamma(0, r)$ and the complementary error function erfc$(r)$ obtained as the results of integration in Eq.(\ref{35}).

It is easy to check that the asymptotic logarithmic derivatives (\ref{36}) satisfy
the inequality (\ref{34c}). 


The substitution
\begin{equation}
\label{37}
y_l(r)= \mathcal{K}(r) \cot \eta(r)
\end{equation}
enables us to transform Eq.(\ref{31}) into the following equation for the phase function $\eta(r)$:
\begin{equation}
\label{38}
\eta'(r)= \dfrac{\mathcal{K}'(r)}{2\mathcal{K}(r)}\sin 2\eta(r)+\mathcal{K}(r)\cos^2\eta(r)-
\dfrac{U(r)}{\mathcal{K}(r)}\sin^2 \eta(r).
\end{equation}
The stabilizing function $\mathcal{K}(r)$ can be chosen in a sufficiently arbitrary manner.
The simplest choice is $\mathcal{K}(r)=1$. In this case, the boundary conditions (\ref{32}) and (\ref{35}) define the following asymptotic condition for $\eta(r)$:
\begin{equation}
\label{38a}
\eta(\infty)=-\pi/2+n\pi .~~~~~~~~~~~~~~~~~~(n=1,2,...)
\end{equation}
Any function $\mathcal{K}(r)$ that preserves the limit, $lim_{r\rightarrow \infty}\left[ y(r)/\mathcal{K}(r)\right] =0$
provides the asymptotic behavior (\ref{38a}).
It was established (at least  numerically) that the more precise results are provided by stabilizing functions of the form:
\begin{equation}
\label{39}
\mathcal{K}_l(r)=\sqrt{U_l(r)},
\end{equation}
where $U_l(r)$ governs the behavior of $U(r)$ both near the origin and at infinity, that is
\begin{equation}
\label{40}
U_l(r)=
\begin{cases}
~ \beta v(r),~~~~~~~~~~~~~~~~~~~~~~~~~~~~~~~~~~~~~~l=0 \\
~l(l+1)r^{-2},~~~~~~~~~~~~~~~~~~~~~~~~~~~~~~~~l>0
\end{cases}.~~~~~~~~~~~~
\end{equation}
Use of the stabilizing functions (\ref{39})-(\ref{40}) enables us to replace Eq.(\ref{38}) with two simpler equations:
\begin{equation}
\label{41}
~~~~~~~\eta'(r)=\dfrac{U_0'(r)}{4U_0(r)}\sin 2\eta(r)+
\sqrt{U_0(r)},~~~~~~~~~~~~~~~~~~~~~~~~~~~~~~~~~~~(l=0)
\end{equation}
\begin{equation}
\label{42}
\eta'(r)=\dfrac{1}{r}\left[\sqrt{l(l+1)}\cos 2\eta(r)-\dfrac{1}{2}\sin2\eta(r) \right] +
\dfrac{r U_0(r)}{\sqrt{l(l+1)}}\sin^2 \eta(r).~~~~~~~~~(l>0)
\end{equation}
Substituting expressions (\ref{39})-(\ref{40}) for $l>0$ into definition
(\ref{37}) with the asymptotic form (\ref{32}), one obtains: $\cot \eta(\infty)=-\sqrt{l/(l+1)}$. Thus, the critical parameters $\beta_n$ must provide the following asymptotic behavior for the function $\eta(r)$:
\begin{equation}
\label{43}
\overset{\sim}{F}_l(\beta_n)\equiv \eta(\infty)\arrowvert_{\beta=\beta_n}=\delta_l-\dfrac{\pi}{2}+n \pi, ~~~~~~~~~~~(n=1,2,...)
\end{equation}
with
\begin{equation}
\label{44}
\delta_l=\arctan\sqrt{\dfrac{l}{l+1}}.
\end{equation}
Eq.(\ref{43}) was derived for $l>0$. It describes the asymptotic behavior of the solutions
of Eq.(\ref{42}). 
However, it is clear that the stabilizing function $\mathcal{K}_0(r)=\sqrt{U_0(r)}$ preserves the correctness of the asymptotic formula (\ref{38a}) for the logarithmic derivatives $y_0(r)$
with the asymptotic behavior defined by Eq.(\ref{35}).
Therefore, condition (\ref{43})-(\ref{44}) with $l=0$ can be used also for the asymptotic approximation of solutions of Eq.(\ref{41}).

It is seen from Eq.(\ref{41}) that the stabilizing function $\mathcal{K}_0(r)=\sqrt{U_0(r)}$ can be applicable
only in case of its nodeless character, otherwise, the simplest choice is $\mathcal{K}_0(r)=1$.

The technique described above in this section is sufficient for presentation of the \textbf{second
method} for calculating the critical parameters of central potentials of the from (\ref{3}).
However, we would like to make some additional remarks that can be useful.

Eqs.(\ref{41}), (\ref{42}), along with the boundary condition (\ref{43})-(\ref{44}), provide a stable and accurate solution to the problem of critical potentials for both small and large values of $l$.

\begin{figure}
\begin{center}
\caption{Function $\overset{\sim}{F}_7(\beta)=\lim_{r\rightarrow\infty}\eta(r)$ for the Gaussian potential ($l=7$). The abscissas of the points of intersection of  $t=\overset{\sim}{F}_7(\beta)$ with lines $t=\delta_l-\pi/2+\pi n$ give the desired critical parameters $\beta=\beta_n$.} 
\epsfxsize=15cm\epsfbox{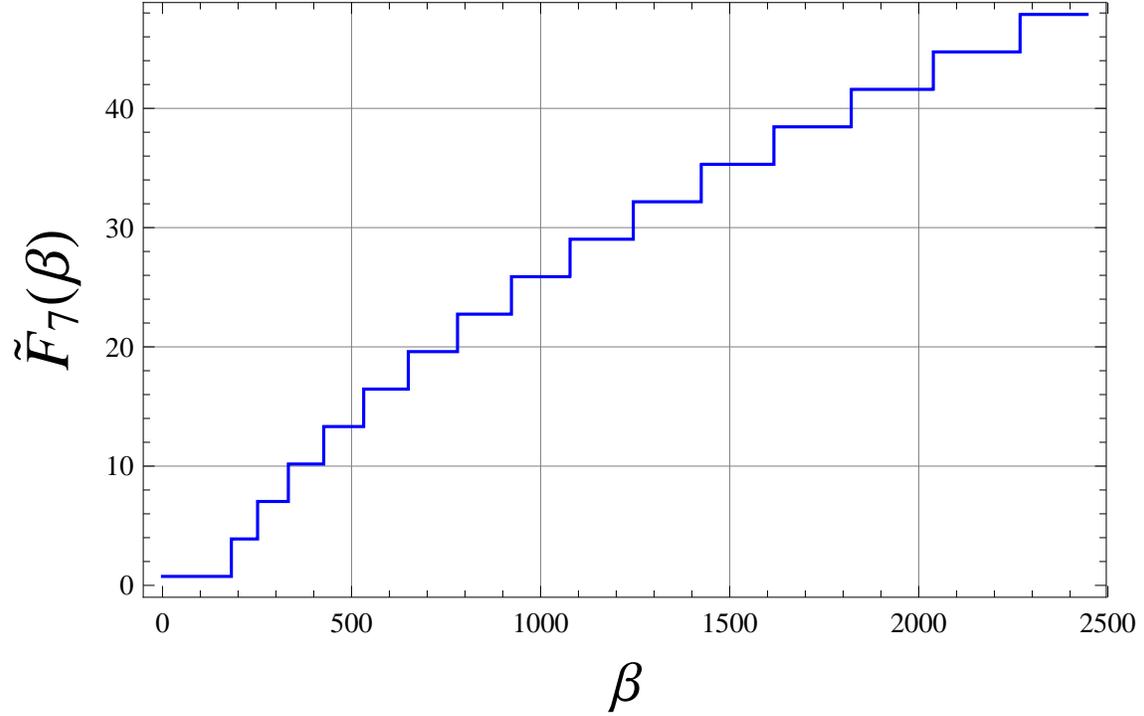} 
\label{F2}
\end{center}
\end{figure}

A typical graph of function $t=\overset{\sim}{F}_l(\beta)$ has a staircase form. It is presented in Fig.\ref{F2}. The abscissas of the points of the staircase function intersections with lines $t=\delta_l-\pi/2+\pi n$ give the desired critical parameters $\beta=\beta_n$.

As an additional useful information it can be shown that
\begin{equation}
\label{45}
\overset{\sim}{F}_l(0)=\delta_l.
\end{equation}
This is because of the following.
Setting $\beta=0$, the second term disappears from the rhs of Eq.(\ref{41}).
The analytic solution to the resultant equation has a form:
\begin{equation}
\label{46}
\eta(r)\underset{\beta\rightarrow 0}{=}\arctan\left(  C\sqrt{U_0(r)}\right) .
\end{equation}
Condition (\ref{1}) thus provides $\eta(\infty)=0$ for $\beta=0$ and $l=0$ according to Eq.(\ref{46}).

Putting $\beta=0$ in Eq.(\ref{42}), the latter loses the term with $U_0(r)$. The resultant equation has analytic solution of the form:
\begin{equation}
\label{47}
\eta(r)\underset{\beta\rightarrow 0}{=}\arctan\dfrac{(2l+1)
\tanh\left( \dfrac{(2l+1)}{2}\ln r+C\right)-1 }{2\sqrt{l(l+1)}}.
\end{equation}
For arbitrary finite real $C$, $tanh$ presented in Eq.(\ref{47}) approaches 1 as $r\rightarrow\infty$.
The resultant expression thus reduces to $\delta_l$.
\section{The first and second order WKB approximations}
\label{sec:wkb}
In this section we apply the first-order and the second-order WKB approaches  
to calculating the critical parameters of central potentials. Unlike the exact methods 
presented earlier these methods are certainly approximate but they are also much simpler.

Specific modification of the first-order WKB approach presented below cannot provide
results of high precision. However, its accuracy grows rapidly with
increasing the number $n$ of bound states. It is important to note that this method can be applied to transition states with any orbital quantum number $l\geq0$. 
 
The accuracy of the second-order WKB method grows with increasing $l$. 
For example, the relative error for $l>10$ can be less than $10^{-5}$.
This enables one to test the results obtained by the phase kind method (Sec.\ref{sec:riccati}) for large $l$. On the other hand, this method can be used by itself if a very high accuracy is not needed. The disadvantage of this method is its inapplicability
for the transition $S$-states ($l=0$).

For $E=0$ the presence of the centrifugal term in the effective
potential $U(r)$ ensures the existence of two turning points for attractive potentials of the form (\ref{3}).
In the second-order approximation \cite{TNI} the WKB quantization condition
\cite{KB1,KB2,KHA}, as applied to our consideration, can be written as
\begin{equation}\label{WKB}
  S_0+S_2=(n-\dfrac{1}{2})\pi,~~~~~~~~~~~~~~~~~~~(n=1,2,...)
\end{equation}
where $n$ is the number of bound states as in the previous sections.
The term
\begin{equation}
\label{49}
S_0=\int_{r_1}^{r_2}\sqrt{-U(r)}dr
\end{equation}
together with the term $S_1=-\pi/2$ correspond to the first order WKB-approximation (for two turning points).
The turning points $r_1$ and $r_2$ are the roots of the equation
\begin{equation}\label{50}
\dfrac{l(l+1)}{r_i^2}=\beta v(r_i),~~~~~~~~~~~~~~~~~~~(i=1,2)
\end{equation}
where the function $v(r)$ is defined by Eqs.(\ref{3}), (\ref{9a}).
The second-order correction takes the form \cite{TNI}
\begin{equation}\label{51}
S_2=\lim_{\mu\rightarrow+0}\left( \dfrac{1}{48}\int_{r_1+\mu}^{r_2-\mu}\dfrac{U''(r)}{[-U(r)]^{3/2}}dr-
\dfrac{1}{12\sqrt{\mu}}(b_1|a_1|^{-3/2}+b_2|a_2|^{-3/2})
\right),
\end{equation}
where $a_1,~b_1,~a_2$ and $b_2$ are the expansion coefficients of the effective potential $U(r)$ in the neighborhood of the turning points. That is
\begin{equation}
\label{52}
U(r_i+\delta r)=a_i\delta r+b_i (\delta r)^2+...
\end{equation}
where from Eq.(\ref{50}) it can be seen that
\begin{eqnarray}
\label{53}
a_i=-\beta\left[\dfrac{2}{r_i}v(r_i)+v'(r_i)\right], ~~~\nonumber~~~~~~~~~~~~~~\\
b_i=\dfrac{1}{2}\beta\left[\dfrac{6}{r_i^2}v(r_i)-v''(r_i) \right].
~~~~~~~~~~~~~~~
\end{eqnarray}
Notice that the small magnitude of the second-order correction (\ref{51}) 
results from the difference of two large terms. Therefore, both these
terms must be calculated with high accuracy. Nevertheless, we would like to
emphasize that this correction increases the accuracy by two to three
orders of magnitude. 

A modification of the first-order WKB method, 
where the centrifugal potential is excluded from the
quasiclassical momentum \cite{KMP}, can be applied for calculating the critical parameters under
consideration. According to this approach, the quantization condition for the transition states ($E=0$)
reduces to:
\begin{equation}
\label{54}
\int_0^\infty\sqrt{\beta v(r)}dr=\pi(n+\gamma_{l,q}),
\end{equation}  
where
\begin{equation}
\label{55}
\gamma_{l,q}=
\begin{cases}
~ \dfrac{2l-1}{4},~~~~~~~~~~~~~~~~~~~~~~~~~~~~~~~~~~~~~~~~~q\leq 0 ~\\
~\dfrac{2l-1+q}{2(2-q)},~~~~~~~~~~~~~~~~~~~~~~~~~~~~~~~~0<q<2
\end{cases}.~~~~~~~~~~~~
\end{equation}
In accordance to its definition (\ref{3a}), the parameter $q$ is ruled by the behavior of the central potential near the origin.
The cut-off potential
(\ref{A1}) presents an exclusion. The exact analytic solution for this case is presented
in the Appendix. 

\section{The critical parameter asymptotics }
\label{sec:asymptotics}

It was shown in the previous sections that in general the critical parameters for 
central potentials can be calculated numerically. For a few special cases, presented
in the Appendix, one can deduce the analytical results.

In this section it will be shown that it is possible to derive the analytical expressions
for the asymptotic form of the critical parameters $\beta\equiv\beta_{n,l}$. In so doing, asymptotic implies a situation where either the number of bound states $n$ approaches
infinity for a given finite $l$, 
or the orbital angular momentum quantum number $l$ goes to infinity for a given finite $n$.
The first order WKB approach is applied to solve this problem.

First, let us consider the case of finite $n$ where $l\rightarrow\infty$.

Numerical calculations demonstrate that the distance between turning points
$|r_1-r_2|$ reaches zero as $l$ approaches infinity. This result can be explained and supported
by the following arguments. According to Eqs.(\ref{WKB}), (\ref{49}) the standard Bohr-Sommerfeld
quantization condition for the transition states reads
\begin{equation}
\label{56}
\int_{r_1}^{r_2}\sqrt{-U(r)}dr=(n-\dfrac{1}{2})\pi.
\end{equation}
This means that the effective potential $U(r)$ must be negative
in the range $[r_1,r_2]$. Hence, according to Eq.(\ref{9}) the critical parameter $\beta$ must tend to infinity as $l^2$ or faster for $l\rightarrow \infty$.
The latter in turn implies that the integrand in the LHS of Eq.(\ref{56}) approaches infinity
as $l\rightarrow \infty$, whereas the RHS of Eq.(\ref{56}) remains finite.
The above contradiction can be eliminated only by setting $r_1=r_2$ for the limits of integration, which proves the statement.

It is clear that the point $R_m$ of minimum of the effective potential $U(r)$ is
localized in the region $[r_1,r_2]$, that is
\begin{equation}
\label{57}
 r_1 \leq  R_m \leq r_2.
\end{equation}
Therefore, for $l$ approaches infinity $R_m$ tends to the point where
the turning points $r_1$ and $r_2$ merge.
Hence, in order to calculate $r_m=\lim_{l\rightarrow \infty}R_m$ it is enough to solve the following set of equations:
\begin{equation}
\label{60}
\begin{cases}
~ U(r_m)=0~~~~~~~~~~~~~~~~~~~ \\
~ U'(r_m)=0~~~~~~~~~~~~~~~~~~~~~
\end{cases}.
\end{equation}
Substituting the explicit form (\ref{9}) into (\ref{60}) and eliminating
$\beta$ and $l$, one obtains the following simple equation for $r_m$:
\begin{equation}
\label{61}
2v(r_m)+r_m v'(r_m)=0.
\end{equation}
 Now, any of the two equations (\ref{60}) gives the required asymptotic expression:
\begin{equation}
\label{62}
\beta_{n,l}\underset{l\rightarrow \infty}{\simeq}\dfrac{d_l}{2}l(l+1),
\end{equation}
with
\begin{equation}
\label{63}
d_l=\dfrac{2}{r_m^2v(r_m)}.
\end{equation}

\begin{table}
\begin{center}
\caption{The asymptotic representations of critical parameters $\beta_{n,l}\underset{l\rightarrow \infty}{\simeq}\dfrac{d_l}{2}l(l+1)$ and~\\ $\beta_{n,l}\underset{n\rightarrow \infty}{\simeq}\dfrac{d_n}{2}\left(n+\gamma_{l,q} \right)^2 $. The numerical results for $\Delta_{n,l}=\beta_{n,l+1}-2\beta_{n,l}+\beta_{n,l-1}$ and~\\ $\varLambda_{n,l}=\beta_{n+1,l}-2\beta_{n,l}+\beta_{n-1,l}$ are presented for comparison.}
\begin{tabular}{|c|c|c|c|c|c|c|}
\hline 
$potential$ & exponential & Hulthen & Yukawa & Gaussian & WS$(x_{0}=1)$ & WS$(x_{0}=0.001)$\tabularnewline
\hline
\hline 
$v(r)$ & $e^{-r}$ & $(e^{r}-1)^{-1}$ & $e^{-r}/r$ & $e^{-r^{2}}$ & $(1+x_{0}e^{r})^{-1}$ & $(1+x_{0}e^{r})^{-1}$\tabularnewline
\hline 
$r_{m}$ & $2$ & $1.59362$ & 1 & 1 & $2.21772$ & $6.17241$\tabularnewline
\hline 
$d_{l}$ & $\frac{e^{2}}{2}/3.69453/$ & $3.08828$ & $2e/5.43656/$ & $2e/5.43656/$ & $4.14224$ & $0.077658$\tabularnewline
\hline 
$\Delta_{1,19}$ & $3.69449$ & $3.08826$ & $5.43654$ & $5.43639$ & $4.14220$ & $0.077664$\tabularnewline
\hline 
$\Delta_{5,19}$ & $3.69674$ & $3.08845$ & $5.43698$ & $5.43895$ & $4.14471$ & $0.077846$\tabularnewline
\hline 
$\intop_{0}^{\infty}{\textstyle \sqrt{v(r)}}dr$ & $2$ & $\pi$ & $\sqrt{2\pi}$ & $\sqrt{\pi/2}$ & $2\ln\frac{1+\sqrt{x_{0}+1}}{\sqrt{x_0}}$ & $2\ln\frac{1+\sqrt{x_{0}+1}}{\sqrt{x_0}} $\tabularnewline
\hline 
$d_{n}$ & $\frac{\pi^{2}}{2}/4.93480/$ & $2$ & $\pi/3.14159/$ & $4\pi/12.5664/$ & $6.35257$ & $0.286909$\tabularnewline
\hline 
$\Lambda_{19,0}$ & $4.93480$ & $2$ & $3.14127$ & $12.5694$ & $6.35257$ & $0.286908$\tabularnewline
\hline 
$\Lambda_{19,5}$ & $4.88254$ & $1.98432$ & $3.11141$ & $12.5151$ & $6.29585$ & $0.286015$\tabularnewline
\hline 
$\Lambda_{99,5}$ & $4.930$ & $1.998$ & $3.138$ & $12.561$ & $6.347$ & $0.2868$\tabularnewline
\hline
\end{tabular}
\label{TL}
\end{center}
\end{table}
The second derivatives $d_l=d^2\beta_{n,l}/dl^2$ of the asymptotic critical parameters (\ref{62}) are presented in Table \ref{TL} along with the values of $\Delta_{n,l}=\beta_{n,l+1}-2\beta_{n,l}+\beta_{n,l-1}$ which approximate the 
general second derivative $d^2\beta_{n,l}/dl^2$ numerically.
It is seen from this Table that the values of $d_l$ are very close to the
values of $\Delta_{1,19}$ for the lowest (nodeless) energy states ($n=1$) and $l=19$.
It is worth noting that even though the Yukawa and Gaussian potentials are very different,
their asymptotic behaviors are coincident ($d_l=2e$).

Notice that a set of equations (\ref{60}) for calculating $r_m$ cannot be applied to the cut-off potential (\ref{A1}). 
The correct result (\ref{A14}) can be obtained by setting $r_m=1$, that is by equating
the merging point $r_m$ and the matching point $r=1$.

Now let us consider the case of $n\rightarrow\infty$ ($l$ is finite).
To this end, one can successfully employ the modified WKB method \cite{KMP}
presented in the previous section. According to the authors of  \cite{KMP}, their method
is ``exact in the asymptotic limit $n_r\rightarrow \infty$``, where $n_r=n-1$.
Our numerical results confirm this assertion. Thus, using directly the quantization
condition (\ref{54}), one obtains: 
\begin{equation}
\label{64}
\beta_{n,l}\underset{n\rightarrow \infty}{\simeq}\dfrac{d_n}{2}\left(n+\gamma_{l,q}\right)^2, 
\end{equation}
where
\begin{equation}
\label{65}
d_n=2\left( \dfrac{\pi}{\int_0^\infty \sqrt{v(r)} dr }\right)^2, 
\end{equation}
and the parameter $\gamma_{l,q}$ is defined in Eq.(\ref{55}).

For the FSW-like potential Eq.(\ref{65}) gives $d_n=2\pi^2\left(\frac{2-s}{2} \right)^2$, which
certainly coincides with the analytic solution presented in the $1^{\underline{st}}$ section of the Appendix.
In general, it is easy to check that the asymptotic expressions for the analytic solutions presented in the Appendix coincide with the results of this section.

\newpage
\section{Numerical results}
\label{sec:results}
The second derivatives $d_n=d^2\beta_{n,l}/dn^2$ of the asymptotic critical parameters (\ref{64}) are presented in Table \ref{TL} along with values of $\varLambda_{n,l}=\beta_{n+1,l}-2\beta_{n,l}+\beta_{n-1,l}$ which approximate the general second derivative $d^2\beta_{n,l}/dn^2$ numerically.
It is seen that for the numerical second derivative $\Delta_{n,l}$ the speed $u_l$ of a convergence to the asymptotic value $d_l$ ($l\rightarrow \infty$) depends on the number of bound states $n$. The fastest convergence corresponds to the smallest $n=1$.
A similar situation is observed for the speed $u_n$ of the convergence of $\varLambda_{n,l}$ to $d_n$ ($n\rightarrow \infty$). In this case, the fastest convergence corresponds 
to the smallest $l=0$.
Moreover, Table \ref{TL} shows that $u_l$ far exceeds $u_n$. 

Using the first and second methods described in Sections
\ref{sec:asymptotic} and \ref{sec:riccati}, respectively, we have computed 
the critical parameters for a few widely used central potentials included in the nonrelativistic Schr\"{o}dinger-equation.
The results for the exponential, Hulthen, Yukawa and Gaussian potentials are presented in Tables \ref{T1}-\ref{T4}, respectively.
The critical parameters for the Woods-Saxon potential with $x_0=1$ and $x_0=0.001$ are exhibited in Tables \ref{T5} and \ref{T6}, respectively. The first value of parameter $x_0$ presents
the minimal value of parameter $R=0$. By contrast, the second value of parameter $x_0$
corresponds to the case of a large value of $R/r_0$ corresponding, e.g., to the optical-model calculations \cite{JEU}.

The critical parameters for the exponential, Hulthen and Woods-Saxon potentials
appearing in the Schr\"{o}dinger equation with 
the orbital angular momentum quantum number $l=0$ ($S$-states) can be calculated analytically.
For the exponential and Woods-Saxon potentials the requested solutions ($l=0$)
are proportional to the squares of zeros of the corresponding special functions (see Appendix). It is worth noting that the critical parameters for the Hulthen potential ($l=0$) have an especially simple form. They are equal to $n^2$ where the number of bound
states equals $n$.
All these parameters are displayed in Tables \ref{T1}, \ref{T2}, \ref{T5} and \ref{T6} for convenient comparing with the cases of $l>0$. Notice that the relative difference between the results obtained by the first method and the analytical ones is less than $10^{-14}$.

The critical parameters for the cut-off potential of the form (\ref{A1}) are not numerically
presented here, because there is no problem to provide the proper calculations according to Eq.(\ref{A7}) for any orbital quantum number $l$ and parameter $s$. However, we have performed
the corresponding computations in order to test both the first and second methods. The relative
difference was less than $10^{-12}$ for the second method (Sec.\ref{sec:riccati}) for $l\leq20$.
The first method provides the same accuracy only for small values of $l\leq3$, whereas
for large $l$ this accuracy can be provided only for $l+n\leq20$. One should notice that all computations were performed by means of the simplest Mathematica-7 codes using the
standard (default) working precision. It is possible, of course, to enhance the calculation
accuracy using, e.g., the better working precision.

Due to lack of space we have restricted the results in our Tables to eight significant figures
and values of $l+n\leq16$.

The Yukawa potential is the only one for which we have revealed some earlier results on critical parameters \cite{ROG}. They are presented there as the critical screening length for the one-electron eigenstates which were obtained in frame of standard energy calculations.
Those results are limited by 5 significant figures and $l+n\leq 9$, and coincide
practically with those exhibited here in Table \ref{T5}.

For the Gaussian potential we have found only the results of the binding energy calculations
(see, e.g., Ref.\cite{ASH,LAI,CRA,BES}). These energies were computed for $l+n\leq8$
and were completely consistent with the critical parameters presented here in Table \ref{T6}.

\section{Conclusions}
\label{sec:conclusions}
In conclusion, we would like to emphasize that the critical parameters are not of some particular character.

First, they present some \textit{universal characteristics} of central potentials.

Second, using the tables of these parameters, one can answer the following questions:

1) What is the number $n$ of bound states for the given central potential and given orbital quantum number $l$.

2) What is the maximum value of $l$ which can provide a bound state for the given central potential, or vice versa, what is the minimum critical parameter which can provide a bound state with a given $l$ for the given central potential.

3) What is the mutual arrangement (order) of the energy levels $E_{n,l}$ (characterized by the quantum numbers $n$ and $l$) for the given form of central potential.

It is clear that the binding energy $E_{n,l}$ rises as the number $n$ of bound states or 
the orbital angular momentum quantum number $l$ increases.
Tables \ref{T1}-\ref{T6} show that the critical parameters $\beta_{n,l}$
exhibit the same properties in respect to the numbers $\{n,l\}$.
It is important to realize that for any two sets $\{n_1,l_1\}$ and $\{n_2,l_2\}$
it follows from inequality
\begin{equation}
\label{90}
\beta_{n_1,l_1}>\beta_{n_2,l_2}
\end{equation}
for the given central potential that
\begin{equation}
\label{91}
E_{n_1,l_1}>E_{n_2,l_2}.
\end{equation}
Therefore, from the presented Tables we can deduce
the following important properties of the discrete energy spectrum
of the considered central potentials:
\begin{eqnarray}
\label{100}
E_{n,l}>E_{n+1,l-1}~~~~~~~~~~~~~~~~~~for~the~Hulthen~and~Yukawa~potentials,
~~~~~~~~~~~~~~~~~~~~~~~\nonumber~~~~~~~~~~\\
E_{n,l}<E_{n+1,l-1}~~~~~~~for~the~exponential,~Gaussian~and~Woods-Saxon~potentials.
~~~\nonumber~~~~~~~~~~~~\\
\end{eqnarray}
This probably relates to the fact that the Hulthen and the Yukawa potentials
are singular at the origin.

It was established that the leading terms of the asymptotic expansions of the critical parameters $\beta_{n,l}$ have the following forms:
\begin{equation}
\label{102}
\beta_{n,l}\simeq
\begin{cases}
~ a_ll^2~~~~~~~~~~~~~~~~~~~~~~~~~~~~~~~~~ l\rightarrow \infty~\\
~ a_nn^2~~~~~~~~~~~~~~~~~~~~~~~~~~~~~~~~n\rightarrow \infty~
\end{cases},~~~~~~~~~~~~
\end{equation}
where the general analytic expressions for the factors $a_l\equiv d_l/2$ and $a_n\equiv d_n/2$ are presented
in section \ref{sec:asymptotics}. The first of the relationships (\ref{102}) is valid
for a finite number $n$ of bound states, whereas the second one is valid for a finite
orbital angular momentum quantum number $l$.

\section{Acknowledgments}

The authors thank Nir Nevo Dinur for useful conversations.

\newpage

\appendix

\section{}

\subsection{Finite square well like potential}

Let us examine potential of the form:
\begin{equation}
\label{A1}
V(r)=
\begin{cases}
~ -\dfrac{g}{r^s},~~~~~~~~~~~~~~~~~~~r\leq r_0 \\
~~~~~0,~~~~~~~~~~~~~~~~~~~~~r>r_0
\end{cases}.~~~~~~~~~~~~(s<2)
\end{equation}
For $s=0$ one obtains the potential which is widely known as the finite square well.
Eq.(\ref{8}) with the potential (\ref{A1}) reduces to the form:

\begin{equation}
\label{A2}
\chi''(r)=\left[ -\dfrac{\beta}{r^s}H(1-r)+\dfrac{l(l+1)}{r^2}\right] \chi(r),
\end{equation}
where $H(x)$ is the Heaviside step function. Parameter $\beta$ is defined by Eq.(\ref{10}).
Particular solution satisfying the asymptotic condition (\ref{20}) and vanishing at the origin has a form:
\begin{equation}
\label{A3}
\chi(r)=
\begin{cases}
~ A \sqrt{r}~J_{\alpha}\left( \dfrac{\sqrt{\beta}}{\nu}r^{\nu}\right) ,~~~~~~~~~~~~~~~~~~~r\leq 1 \\
~~~~~B r^{-l}~~~~~~~~~~~~~~~~~~~~~~~~~~~~~~~~~~r>1
\end{cases},
\end{equation}
with
\begin{equation}
\label{A4}
\nu=\frac{2-s}{2},~~~~~~~~~~~~~~~~~~\alpha=\frac{2l+1}{2-s}.
\end{equation}
Here $J_\alpha(z)$ is the Bessel function of the first kind, whereas $A$ and $B$ are arbitrary constants.
Matching the logarithmic derivatives of solutions (\ref{A3}) at the point $r=1$, one obtains the following equation for $\beta$:
\begin{equation}
\label{A5}
\sqrt{\beta}\left[(l+1)J_{\alpha-1}\left(\frac{\sqrt{\beta}}{\nu}\right)
-lJ_{\alpha+1}\left(\frac{\sqrt{\beta}}{\nu}\right)\right]=
-l(2l+1)J_\alpha \left(\frac{\sqrt{\beta}}{\nu}\right).
\end{equation}
Using the properties of the Bessel functions \cite{ABR}, one can reduce Eq.(\ref{A5}) to the following simplest equation:
\begin{equation}
\label{A6}
J_{\frac{2l-1+s}{2-s}}\left(\dfrac{2\sqrt{\beta}}{2-s} \right)=0.
\end{equation}
The explicit solution of Eq.(\ref{A6}) has a form:
\begin{equation}
\label{A7}
\beta_n=\left[\left(\frac{2-s}{2}\right)j_{\frac{2l-1+s}{2-s},~n}\right]^2,~~~~~~~~~~~~~~(n=1,2,...)
\end{equation}
where $j_{\mu,n}$ presents  the $n^{th}$ positive zero of the Bessel function
$J_\mu(z)$.

From expansion (9.5.12)\cite{ABR} for large zeros one has:
\begin{equation}
\label{A7a}
j_{\nu,n}\simeq\left(n+\dfrac{\nu}{2}-\dfrac{1}{4} \right)\pi-\dfrac{4\nu^2-1}{8\left(n+\dfrac{\nu}{2}-\dfrac{1}{4} \right)\pi}-
\dfrac{4(4\nu^2-1)(28\nu^2-31)}{192\left(n+\dfrac{\nu}{2}-\dfrac{1}{4} \right)^3\pi^3}-... 
\end{equation}
Putting
\begin{equation}
\label{A7b}
\nu\equiv\nu_{l,s}=\dfrac{2}{2-s}l-\dfrac{1-s}{2-s},
\end{equation}
and using Eq.(\ref{A7}), one obtains the following asymptotic expression for the critical
parameters corresponding to potential (\ref{A1}):
\begin{equation}
\label{A7c}
\beta_n\underset{n\rightarrow \infty}{\simeq}\left( \dfrac{2-s}{2}\right)^2\pi^2n^2. 
\end{equation}
This yields, in particular:
\begin{equation}
\label{A7d}
\dfrac{d^2}{d n^2}\beta_n(s=0)\underset{n\rightarrow \infty}{\simeq}2\pi^2,
~~~~~~~\dfrac{d^2}{d n^2}\beta_n(s=1)\underset{n\rightarrow \infty}{\simeq}\pi^2/2.
\end{equation}
From expansion for zeros of the Bessel functions of large order \cite{OLV} (see, also (9.5.14)\cite{ABR}), one has:
\begin{equation}
\label{A7e}
j_{\nu,n}=\nu\left(1+\sum_{k=1}^\infty \alpha_{k,n}\nu^{-\dfrac{2k}{3}}\right). 
\end{equation}
The first coefficients $\alpha_{k,n}$ for $\{k,n\}\leq 5$ can be found in \cite{OLV}.
Thus, using Eq.(\ref{A7b}), one obtains for large enough $l$:
\begin{equation}
\label{A7f}
\beta_n\equiv\beta_{n,l}=\left(l-\dfrac{1-s}{2} \right)^2
\left[1+\sum_{k=1}^\infty \alpha_{k,n}\left(\dfrac{2l-1+s}{2-s} \right)^{-\dfrac{2k}{3}} \right]^2.  
\end{equation}
It is seen that the leading term of the asymptotic ($l\rightarrow\infty$) expansion of the critical parameter is given by
\begin{equation}
\label{A7g}
\beta_{n,l}\underset{l\rightarrow \infty}{\simeq}l^2. 
\end{equation}

\subsection{Exponential potential}

The exponential potential has a form:
\begin{equation}
\label{A8}
V(r)=-g~exp\left(-\frac{r}{r_0}\right).
\end{equation}
Eq.(\ref{8}) with the potential (\ref{A8}) reads
\begin{equation}
\label{A9}
\chi''(r)=\left[ -\beta~exp(-r)+\dfrac{l(l+1)}{r^2}\right] \chi(r).
\end{equation}
Parameter $\beta$ is defined here by Eq.(\ref{10}) with $s=0$.
Eq.(\ref{A9}) has an analytical solution only for the case of $l=0$.
Such a particular solution satisfying the boundary condition (\ref{14}) has a form:
\begin{equation}
\label{A10}
\chi(r)=A\left[J_0(q)Y_0\left(qe^{-\frac{r}{2}} \right)-Y_0(q)J_0\left(qe^{-\frac{r}{2}} \right) \right],
\end{equation}
where $q=2\sqrt{\beta}$, whereas $J_0(z)$ and $Y_0(z)$ are the Bessel functions of the
first and second kind, respectively.
The argument of the Bessel functions in Eq.(\ref{A10}) achieves zero as $r\rightarrow \infty$. Therefore, using series expansion for the Bessel functions \cite{ABR}
\begin{equation}
\label{A11}
J_0\left(qe^{-\frac{r}{2}} \right)\underset{r\rightarrow \infty}{\simeq}1-\dfrac{q^2}{4}e^{-r},~~~~~~~~~
 Y_0\left(qe^{-\frac{r}{2}} \right)\underset{r\rightarrow \infty}{\simeq}-\frac{r}{\pi}
\end{equation}
one should put
\begin{equation}
\label{A12}
J_0(q)=0,
\end{equation}
in order solution (\ref{A10}) satisfies the asymptotic boundary condition (\ref{20}).
Taking into account condition (\ref{A12}) for the transition state, and the first of expansions (\ref{A11}) one can write for the asymptotic behavior of the solution (\ref{A10}):
\begin{equation}
\label{A12a}
\chi(r)\underset{r\rightarrow \infty}{\simeq}-AY_0(2\sqrt{\beta})\left( 1-\beta e^{-r}\right) .
\end{equation} 
From Eq.(\ref{A12}) one obtains for the critical parameters: 
\begin{equation}
\label{A13}
\beta_n=\frac{1}{4}j_{0,n}^2~,~~~~~~~~~~~~~~~~~~~~~~~~~~~~~(n=1,2,...)
\end{equation}
where $j_{0,n}$ presents  the $n^{th}$ positive zero of the Bessel function $J_0(z)$.

Putting $\nu=0$ in expansion (\ref{A7a}) for large zeros, one obtains:
\begin{equation}
\label{A13a}
j_{0,n}\simeq\left(n-\dfrac{1}{4} \right)\pi+\dfrac{1}{2(4n-1)\pi}-\dfrac{124}{3(4n-1)^3\pi^3}+... 
\end{equation}
This yields
\begin{equation}
\label{A13b}
\beta_n\underset{n\rightarrow \infty}{\simeq}\dfrac{\pi^2}{4}\left(n-\dfrac{1}{4} \right) ^2. 
\end{equation}
The value of second derivative $\lim_{n\rightarrow \infty}d^2\beta_n/dn^2=\pi^2/2$
coincides with the corresponding value (\ref{A7d}) for the FSW-like potential with $s=1$.

\subsection{Hulthen potential}

For the Hulthen potential
\begin{equation}
\label{A14}
V(r)=-\frac{g~ e^{-\frac{r}{r_0}}}{1-e^{-\frac{r}{r_0}}},
\end{equation}
Eq.(\ref{8}) becomes
\begin{equation}
\label{A15}
\chi''(r)=\left[ -\frac{\beta}{e^r-1}+\dfrac{l(l+1)}{r^2}\right] \chi(r).
\end{equation}
For this differential equation one can obtain a general analytic solution of the form
\begin{equation}
\label{A16}
\chi(r)\equiv \phi(x)=A~ {x^{-\alpha}}~_2F_1(-\alpha,-\alpha;1-2\alpha;x)+
B~{x^{\alpha}}~_2F_1(\alpha,\alpha;1+2\alpha;x)
\end{equation}
with
\begin{equation}
\label{A17}
\alpha=\sqrt{\beta},~~~~~~~~~~~~~~~~~~~ x=e^r,
\end{equation}
only for $S$-states ($l=0$). Here $~_2 F_1(a,b;c;z)$ is the Gauss hypergeometric function, $A$ and $B$ are arbitrary constants.

Formulas (15.1.20) and (6.1.18) \cite{ABR} yield:
\begin{equation}
\label{A18}
~_2 F_1(a,a;1+2a;1)=\frac{4^a~\Gamma(a+1/2)}{\sqrt{\pi}~\Gamma(a+1)},
\end{equation}
where $\Gamma(z)$ denotes Euler's gamma function.
The latter representation enables us to obtain the vanishing at the origin
($r\rightarrow 0 \Rightarrow  x\rightarrow 1$  ) solution in the form:
\begin{eqnarray}
\label{A19}
\chi(r)\equiv\varphi(x)= C\left[ \left(\frac{4}{x}\right)^\alpha\Gamma(1-\alpha)\Gamma\left(\frac{1}{2}+\alpha\right)
~_2F_1(-\alpha,-\alpha;1-2\alpha;x)-\right.
~~~~~~~~~~~~~~~~~~~~~~\nonumber~~~~~~~\\
\left.\left(\frac{x}{4}\right)^\alpha\Gamma(1+\alpha)\Gamma\left(\frac{1}{2}-\alpha\right)
~_2F_1(\alpha,\alpha;1+2\alpha;x)\right],~~~~~~~~~~~~~~
\end{eqnarray}
with arbitrary constant $C$.
For examination of the asymptotic behavior of solution (\ref{A19}), one can use
formula (15.3.13)\cite{ABR}, which yields:
\begin{equation}
\label{A20}
(-x)^a~_2F_1(a,a;1+2a;x)\underset{x\rightarrow \infty}{\simeq}
\frac{2\Gamma(2a)}{\Gamma^2(a)}\ln(-x).
\end{equation}
Inserting the asymptotic representation (\ref{A20}) for $a=\alpha$ and $a=-\alpha$ into the rhs of Eq.(\ref{A19}), one obtains after some transformations:
\begin{equation}
\label{A21}
\chi(r)\underset{r\rightarrow \infty}{\simeq}-2C\alpha\sqrt{\pi}(i\pi+r).
\end{equation}
On the other hand, series expansion of solution (\ref{A19}) near $x=1$ ($r\rightarrow0$) yields:
\begin{equation}
\label{A22}
\chi(r)\underset{r\rightarrow 0}{\simeq}-2C\alpha^2 \pi^{3/2}\csc(\pi \alpha)r.
\end{equation}
Putting
\begin{equation}
\label{A23}
C=-[2\alpha^2\pi^{3/2}csc(\pi \alpha)]^{-1}
\end{equation}
one can get rid of $\alpha$-dependence for the leading term of the $\chi(r)$ series expansion. Substituting expression (\ref{A23}) into the asymptotic representation (\ref{A21}), one finally obtains:
\begin{equation}
\label{A24}
\chi(r)\underset{r\rightarrow \infty}{\simeq}\frac{\sin(\pi \alpha)}{\pi \alpha}(i\pi+r).
\end{equation}
Thus, in order to satisfy asymptotic condition (\ref{20}), one should put:
\begin{equation}
\label{A25}
\sin(\pi \alpha)=0.~~~~~~~~~~~~~~(\alpha\neq 0)
\end{equation}
The roots of Eq.(\ref{A25}) are the integers, that is $\alpha_n=n$.
Thus, from definition (\ref{A17}) one obtains that the critical parameters for the Hulthen potential can be determined from the simplest relation:
\begin{equation}
\label{A26}
\frac{2 m g r_0^2}{\hbar^2}=n^2,~~~~~~~~~~~~~~~~~~~~~~~(n=1,2,...)
\end{equation}
where $n$ is a number of $S$-bound states.

\subsection{Woods-Saxon potential}

Finally, let us examine the Woods-Saxon potential
\begin{equation}
\label{A27}
V(r)=-\dfrac{g}{1+exp(\dfrac{r-R}{r_0})}~~~~~~~~~~~(R>0)
\end{equation}
which is the most complicated one. For this case Eq.(\ref{8}) takes a form:
\begin{equation}
\label{A28}
\chi''(r)=\left[ -\frac{\beta}{1+x_0e^r}+\dfrac{l(l+1)}{r^2}\right] \chi(r),
\end{equation}
where $\beta$ is defined by Eq.(\ref{10}) with $s=0$, whereas $x_0=exp(-R/r_0)$.
Eq.(\ref{A28}) admits analytical solution only for the case of $l=0$. Introducing a new variable
$x=x_0e^r$ and a new parameter $\alpha=\sqrt{\beta}$, one obtains a new differential equation
\begin{equation}
\label{A29}
x^2\varphi''(x)+x \varphi'(x)+\dfrac{\alpha^2}{1+x}\varphi(x)=0
\end{equation}
for function $\varphi(x)\equiv\chi(r)$. The vanishing at the origin solution of Eq.(\ref{A29}) has a form:
\begin{equation}
\label{A30}
\varphi(x)=C\left[F(-\alpha,x_0)F(\alpha,x)-F(\alpha,x_0)F(-\alpha,x) \right],
\end{equation}
where
\begin{equation}
\label{A31}
F(\alpha,x)=x^{i\alpha}~_2F_1(i\alpha,i\alpha;1+2i\alpha;-x).
\end{equation}
Here $~_2F_1(a,b;c;z)$ is the Gauss hypergeometric function, $C$ is arbitrary constant.
For considered case, formula (15.3.13) \cite{ABR} yields
\begin{equation}
\label{A32}
(x)^b~_2F_1(b,b;1+2b;-x)\underset{x\rightarrow \infty}{\simeq}
\frac{2\Gamma(2b)}{\Gamma^2(b)}\ln(x).
\end{equation}
Inserting the latter representation into solution (\ref{A30}) and returning to
the initial variable $r$, one obtains:
\begin{eqnarray}
\label{A33}
\chi(r)\underset{r\rightarrow \infty}{\simeq}2C(r+\ln x_0)
\left[\dfrac{F(-\alpha,x_0)\Gamma(2i\alpha)}{\Gamma^2(i\alpha)}-
\dfrac{F(\alpha,x_0)\Gamma(-2i\alpha)}{\Gamma^2(-i\alpha)} \right]=
~~~~~~~~~~~~~~~~~~~~~~\nonumber~~~~~~~\\
2C(r+\ln x_0)i\alpha~_2F_1\left(-i\alpha,i\alpha;1;-\dfrac{1}{x_0} \right). ~~~~~~~~~~~~~~~~~~~~~~~~
\end{eqnarray}
Thus, to satisfy the condition (\ref{20}) for the transitional states, one should put
\begin{equation}
\label{A34}
~_2F_1\left(-i\alpha,i\alpha;1;-\dfrac{1}{x_0} \right)=0.
\end{equation}
The roots $\beta_n=\alpha_n^2$ of the latter transcendental equation present
the desired critical parameters for a given $x_0$.
It is worth noting that Eq.(\ref{A34}) can be simplified if one uses the following relationships
between the Gauss hypergeometric functions, the Jacobi functions $P_\nu^{(a,b)}$ and
the Legendre functions $P_\nu$:
\begin{eqnarray}
\label{A35}
~_2F_1\left(-i\alpha,i\alpha;1;-z\right) =P_{i\alpha}^{(0,-1)}(2z+1)=
(1+z)P_{i\alpha-1}^{(0,1)}(2z+1)=Re\left[P_{i\alpha}(2z+1) \right]~~~\nonumber~~~\\
(\alpha>0,~z>0).~~~~~~~~~
\end{eqnarray}
According to the asymptotic expansion for the Legendre function of imaginary degree
(see, Eq.(3.2) \cite{MAL}), one has
\begin{equation}
\label{A36}
Re\left[P_{i\alpha}(cosh~t) \right]\simeq\dfrac{1}{\sqrt{2}}\sum_{k=0}^N
(2k-1)!!a_k(t)\left(-\dfrac{t}{\alpha} \right)^k J_k(\alpha t)+O\left(\alpha^{-N-1} \right),  
\end{equation}
where
\begin{equation}
\label{A37}
a_0(t)=\sqrt{t~coth\left(\dfrac{t}{2}\right)},~~
a_1(t)=\dfrac{a_0(t)}{8t}\left( \dfrac{cosh~t-2}{sinh ~t}+\dfrac{1}{t}\right). 
\end{equation}
In zero approximation for large enough $\alpha$, one can put $N=0$, whence
\begin{equation}
\label{A38}
Re\left[P_{i\alpha}(cosh~t) \right]\underset{\alpha\rightarrow \infty}{\simeq}
\dfrac{1}{\sqrt{2}}\sqrt{t~coth\left(\dfrac{t}{2} \right) }J_0(\alpha t).
\end{equation}
Thus, roots of Eq.(\ref{A34}) for large enough $\alpha=\sqrt{\beta}$ are very close
to zeros of the Bessel function $J_0(\alpha t)$, where
\begin{equation}
\label{A39}
t=arccosh\left( \dfrac{2}{x_0}+1\right)=2 arcsinh\left(\dfrac{1}{\sqrt{x_0}} \right)=
2\ln \left(\dfrac{1+\sqrt{x_0+1}}{\sqrt{x_0}} \right).
\end{equation}
In terms of zeros $j_{0,n}$ of the Bessel functions, one can then write down:
\begin{equation}
\label{A40}
j_{0,n}\simeq2\sqrt{\beta}\ln \left(\dfrac{1+\sqrt{x_0+1}}{\sqrt{x_0}} \right).
\end{equation}
Taking into account expansion (\ref{A13a}), one obtains finally:
\begin{equation}
\label{A41}
\beta_n\underset{n\rightarrow \infty}{\simeq}\dfrac{\pi^2 }{4\ln^2\left(\dfrac{1+\sqrt{x_0+1}}{\sqrt{x_0}} \right)}\left(n-\dfrac{1}{4} \right)^2.
\end{equation}

\newpage

\begin{table}
\begin{center}
\caption{Critical parameters $\beta=2mgr_0^2\hbar^{-2}$ of the exponential potential $V(r)=-g~\exp(-r/r_0)$.}
\begin{tabular}{|c|ccccccccc|}
\hline
{\footnotesize $n\diagdown l$} & {\footnotesize 0} & {\footnotesize 1} & {\footnotesize 2} & {\footnotesize 3} & {\footnotesize 4} & {\footnotesize 5} & {\footnotesize 6} & {\footnotesize 7} & {\footnotesize 8}\tabularnewline
\hline
{\footnotesize 1} & {\footnotesize 1.4457965} & {\footnotesize 7.0490613} & {\footnotesize 16.312928} & {\footnotesize 29.258323} & {\footnotesize 45.892427} & {\footnotesize 66.218077} & {\footnotesize 90.236557} & {\footnotesize 117.94852} & {\footnotesize 149.35431}\tabularnewline
{\footnotesize 2} & {\footnotesize 7.6178156} & {\footnotesize 16.921126} & {\footnotesize 29.879667} & {\footnotesize 46.518231} & {\footnotesize 66.845206} & {\footnotesize 90.863808} & {\footnotesize 118.57542} & {\footnotesize 149.98069} & {\footnotesize 185.07997}\tabularnewline
{\footnotesize 3} & {\footnotesize 18.721752} & {\footnotesize 31.525958} & {\footnotesize 48.076670} & {\footnotesize 68.345880} & {\footnotesize 92.323606} & {\footnotesize 120.00486} & {\footnotesize 151.38676} & {\footnotesize 186.46751} & {\footnotesize 225.24597}\tabularnewline
{\footnotesize 4} & {\footnotesize 34.760071} & {\footnotesize 50.947660} & {\footnotesize 71.002111} & {\footnotesize 94.837734} & {\footnotesize 122.41832} & {\footnotesize 153.72542} & {\footnotesize 188.74853} & {\footnotesize 227.48127} & {\footnotesize 269.91954}\tabularnewline
{\footnotesize 5} & {\footnotesize 55.733076} & {\footnotesize 75.226301} & {\footnotesize 98.713352} & {\footnotesize 126.05709} & {\footnotesize 157.19351} & {\footnotesize 192.08825} & {\footnotesize 230.72116} & {\footnotesize 273.07970} & {\footnotesize 319.15566}\tabularnewline
{\footnotesize 6} & {\footnotesize 81.640838} & {\footnotesize 104.38402} & {\footnotesize 131.24653} & {\footnotesize 162.04738} & {\footnotesize 196.69590} & {\footnotesize 235.14112} & {\footnotesize 277.35218} & {\footnotesize 323.30940} & {\footnotesize 372.99967}\tabularnewline
{\footnotesize 7} & {\footnotesize 112.48338} & {\footnotesize 138.43438} & {\footnotesize 168.62580} & {\footnotesize 202.83958} & {\footnotesize 240.96044} & {\footnotesize 282.92106} & {\footnotesize 328.67953} & {\footnotesize 378.20847} & {\footnotesize 431.48931}\tabularnewline
{\footnotesize 8} & {\footnotesize 148.26072} & {\footnotesize 177.38629} & {\footnotesize 210.86809} & {\footnotesize 248.45649} & {\footnotesize 290.01388} & {\footnotesize 335.45733} & {\footnotesize 384.73391} & {\footnotesize 437.80834} & {\footnotesize 494.65627}\tabularnewline
\cline{10-10}
{\footnotesize 9} & {\footnotesize 188.97285} & {\footnotesize 221.24597} & {\footnotesize 257.98574} & {\footnotesize 298.91536} & {\footnotesize 343.87710} & {\footnotesize 392.77335} & {\footnotesize 445.54045} & \multicolumn{1}{c|}{{\footnotesize 502.13521}} & \multicolumn{1}{c|}{}\tabularnewline
\cline{9-9}
{\footnotesize 10} & {\footnotesize 234.61978} & {\footnotesize 270.01793} & {\footnotesize 309.98801} & {\footnotesize 354.22957} & {\footnotesize 402.56672} & {\footnotesize 454.88817} & \multicolumn{1}{c|}{{\footnotesize 511.11996}} & \multicolumn{1}{c|}{{\footnotesize 488.5090}} & \multicolumn{1}{c|}{{\footnotesize 7}}\tabularnewline
\cline{8-8}
{\footnotesize 11} & {\footnotesize 285.20151} & {\footnotesize 323.70558} & {\footnotesize 366.88206} & {\footnotesize 414.40969} & {\footnotesize 466.09616} & \multicolumn{1}{c|}{{\footnotesize 521.81745}} & \multicolumn{1}{c}{{\footnotesize 483.54571}} & \multicolumn{1}{c|}{{\footnotesize 426.4139}} & \multicolumn{1}{c|}{{\footnotesize 6}}\tabularnewline
\cline{7-7}
{\footnotesize 12} & {\footnotesize 340.71804} & {\footnotesize 382.31153} & {\footnotesize 428.67350} & {\footnotesize 479.46420} & \multicolumn{1}{c|}{{\footnotesize 534.47640}} & \multicolumn{1}{c}{{\footnotesize 479.64012}} & {\footnotesize 422.43924} & \multicolumn{1}{c|}{{\footnotesize 368.94348}} & \multicolumn{1}{c|}{{\footnotesize 5}}\tabularnewline
\cline{6-6}
{\footnotesize 13} & {\footnotesize 401.16937} & {\footnotesize 445.83785} & {\footnotesize 495.36688} & \multicolumn{1}{c|}{{\footnotesize 549.40007}} & \multicolumn{1}{c}{{\footnotesize 476.68376}} & {\footnotesize 419.44387} & {\footnotesize 365.90248} & \multicolumn{1}{c|}{{\footnotesize 316.06058}} & \multicolumn{1}{c|}{{\footnotesize 4}}\tabularnewline
\cline{5-5}
{\footnotesize 14} & {\footnotesize 466.55550} & {\footnotesize 514.28624} & \multicolumn{1}{c|}{{\footnotesize 566.96589}} & \multicolumn{1}{c}{{\footnotesize 474.58293}} & {\footnotesize 417.32411} & {\footnotesize 363.76075} & {\footnotesize 313.89306} & \multicolumn{1}{c|}{{\footnotesize 267.72133}} & \multicolumn{1}{c|}{{\footnotesize 3}}\tabularnewline
\cline{4-4}
{\footnotesize 15} & {\footnotesize 536.87643} & \multicolumn{1}{c|}{{\footnotesize 587.65804}} & \multicolumn{1}{c}{{\footnotesize 473.25639}} & {\footnotesize 415.99092} & {\footnotesize 362.41992} & {\footnotesize 312.54336} & {\footnotesize 266.36122} & \multicolumn{1}{c|}{{\footnotesize 223.87344}} & \multicolumn{1}{c|}{{\footnotesize 2}}\tabularnewline
\cline{3-3}
{\footnotesize 16} & \multicolumn{1}{c|}{{\footnotesize 612.13217}} & \multicolumn{1}{c}{{\footnotesize 472.63339}} & {\footnotesize 415.36754} & {\footnotesize 361.79612} & {\footnotesize 311.91912} & {\footnotesize 265.736481} & {\footnotesize 223.24818} & \multicolumn{1}{c|}{{\footnotesize 184.45415}} & \multicolumn{1}{c|}{{\footnotesize 1}}\tabularnewline
\cline{3-3}
\cline{1-1} \cline{2-2} \cline{4-4} \cline{5-5} \cline{6-6} \cline{7-7} \cline{8-8} \cline{9-9} \cline{10-10}
\multicolumn{1}{c}{} & \multicolumn{1}{c|}{} & \multicolumn{1}{c}{{\footnotesize 15}} & {\footnotesize 14} & {\footnotesize 13} & {\footnotesize 12} & {\footnotesize 11} & {\footnotesize 10} & \multicolumn{1}{c|}{{\footnotesize 9}} & \multicolumn{1}{c|}{{\footnotesize $l\,\diagdown n$}}\tabularnewline
\cline{3-3} \cline{4-4} \cline{5-5} \cline{6-6} \cline{7-7} \cline{8-8} \cline{9-9} \cline{10-10}
\end{tabular}
\label{T1}
\end{center}
\end{table}

\begin{table}
\begin{center}
\caption{Critical parameters $\beta=2mgr_0^2\hbar^{-2}$ of the Hulthen potential~\\ $V(r)=-g~\exp(-r/r_0)/[1-\exp(-r/r_0)]$.}
\begin{tabular}{|c|ccccccccc|}
\hline
{\footnotesize $n\diagdown l$} & {\footnotesize 0} & {\footnotesize 1} & {\footnotesize 2} & {\footnotesize 3} & {\footnotesize 4} & {\footnotesize 5} & {\footnotesize 6} & {\footnotesize 7} & {\footnotesize 8}\tabularnewline
\hline
{\footnotesize 1} & {\footnotesize 1} & {\footnotesize 5.3059406} & {\footnotesize 12.685368} & {\footnotesize 23.146783} & {\footnotesize 36.693671} & {\footnotesize 53.327421} & {\footnotesize 73.048651} & {\footnotesize 95.857670} & {\footnotesize 121.75465}\tabularnewline
{\footnotesize 2} & {\footnotesize 4} & {\footnotesize 10.724673} & {\footnotesize 20.499398} & {\footnotesize 33.348492} & {\footnotesize 49.279726} & {\footnotesize 68.296123} & {\footnotesize 90.399036} & {\footnotesize 115.58915} & {\footnotesize 143.86682 }\tabularnewline
{\footnotesize 3} & {\footnotesize 9} & {\footnotesize 18.100968} & {\footnotesize 30.253614} & {\footnotesize 45.480804} & {\footnotesize 63.790287} & {\footnotesize 85.185101} & {\footnotesize 109.66659} & {\footnotesize 137.23540} & {\footnotesize 167.89189}\tabularnewline
{\footnotesize 4} & {\footnotesize 16} & {\footnotesize 27.448626} & {\footnotesize 41.962702} & {\footnotesize 59.557425} & {\footnotesize 80.237830} & {\footnotesize 104.00568} & {\footnotesize 130.86162} & {\footnotesize 160.80588} & {\footnotesize 193.83855}\tabularnewline
{\footnotesize 5} & {\footnotesize 25} & {\footnotesize 38.775435} & {\footnotesize 55.636363 } & {\footnotesize 75.588244} & {\footnotesize 98.631880 } & {\footnotesize 124.76686} & {\footnotesize 153.99258} & {\footnotesize 186.30851 } & {\footnotesize 221.71423}\tabularnewline
{\footnotesize 6} & {\footnotesize 36} & {\footnotesize 52.086245} & {\footnotesize 71.281353} & {\footnotesize 93.580632 } & {\footnotesize 118.97988} & {\footnotesize 147.47591} & {\footnotesize 179.06647} & {\footnotesize 213.74998} & {\footnotesize 251.52532}\tabularnewline
{\footnotesize 7} & {\footnotesize 49} & {\footnotesize 67.384298} & {\footnotesize 88.902568} & {\footnotesize 113.54024 } & {\footnotesize 141.28775} & {\footnotesize 172.13879} & {\footnotesize 206.08916} & {\footnotesize 243.13600} & {\footnotesize 283.27733}\tabularnewline
{\footnotesize 8} & {\footnotesize 64} & {\footnotesize 84.671878} & {\footnotesize 108.50368} & {\footnotesize 135.47149} & {\footnotesize 165.56029 } & {\footnotesize 198.76046} & {\footnotesize 235.06562} & {\footnotesize 274.47149} & {\footnotesize 316.97510}\tabularnewline
\cline{10-10}
{\footnotesize 9} & {\footnotesize 81} & {\footnotesize 103.95066} & {\footnotesize 130.08753} & {\footnotesize 159.37792} & {\footnotesize 191.80146} & {\footnotesize 227.34506 } & {\footnotesize 266.00010} & \multicolumn{1}{c|}{{\footnotesize 307.76072}} & \multicolumn{1}{c|}{}\tabularnewline
\cline{9-9}
{\footnotesize 10} & {\footnotesize 100} & {\footnotesize 125.22192 } & {\footnotesize 153.65633} & {\footnotesize 185.26240} & {\footnotesize 220.01452} & {\footnotesize 257.89613} & \multicolumn{1}{c|}{{\footnotesize 298.89626}} & \multicolumn{1}{c|}{{\footnotesize 326.51175}} & \multicolumn{1}{c|}{{\footnotesize 7}}\tabularnewline
\cline{8-8}
{\footnotesize 11} & {\footnotesize 121} & {\footnotesize 148.48665 } & {\footnotesize 179.21189} & {\footnotesize 213.12731} & {\footnotesize 250.20224} & \multicolumn{1}{c|}{{\footnotesize 290.41668 }} & \multicolumn{1}{c}{{\footnotesize 336.34850}} & \multicolumn{1}{c|}{{\footnotesize 292.39168}} & \multicolumn{1}{c|}{{\footnotesize 6}}\tabularnewline
\cline{7-7}
{\footnotesize 12} & {\footnotesize 144} & {\footnotesize 173.74563} & {\footnotesize 206.75567} & {\footnotesize 242.97462} & \multicolumn{1}{c|}{{\footnotesize 282.36696 }} & \multicolumn{1}{c}{{\footnotesize 346.46731 }} & {\footnotesize 301.79385} & \multicolumn{1}{c|}{{\footnotesize 260.20943}} & \multicolumn{1}{c|}{{\footnotesize 5}}\tabularnewline
\cline{6-6}
{\footnotesize 13} & {\footnotesize 169} & {\footnotesize 200.99951} & {\footnotesize 236.28888 } & \multicolumn{1}{c|}{{\footnotesize 274.80601}} & \multicolumn{1}{c}{{\footnotesize 356.85339}} & {\footnotesize 311.46707} & {\footnotesize 269.16916} & \multicolumn{1}{c|}{{\footnotesize 229.95964}} & \multicolumn{1}{c|}{{\footnotesize 4}}\tabularnewline
\cline{5-5}
{\footnotesize 14} & {\footnotesize 196} & {\footnotesize 230.24882} & \multicolumn{1}{c|}{{\footnotesize 267.81254}} & \multicolumn{1}{c}{{\footnotesize 367.49439}} & {\footnotesize 321.39765} & {\footnotesize 278.38907} & {\footnotesize 238.46863} & \multicolumn{1}{c|}{{\footnotesize 201.63626}} & \multicolumn{1}{c|}{{\footnotesize 3}}\tabularnewline
\cline{4-4}
{\footnotesize 15} & {\footnotesize 225} & \multicolumn{1}{c|}{{\footnotesize 261.49399 }} & \multicolumn{1}{c}{{\footnotesize 378.38000}} & {\footnotesize 331.57422} & {\footnotesize 287.85660} & {\footnotesize 247.22711} & {\footnotesize 209.68569} & \multicolumn{1}{c|}{{\footnotesize 175.23229}} & \multicolumn{1}{c|}{{\footnotesize 2}}\tabularnewline
\cline{3-3}
{\footnotesize 16} & \multicolumn{1}{c|}{{\footnotesize 256}} & \multicolumn{1}{c}{{\footnotesize 389.50148}} & {\footnotesize 341.98731} & {\footnotesize 297.56137} & {\footnotesize 256.22366} & {\footnotesize 217.97415} & {\footnotesize 182.81284} & \multicolumn{1}{c|}{{\footnotesize 150.73968}} & \multicolumn{1}{c|}{{\footnotesize 1}}\tabularnewline
\cline{3-3}
\cline{1-1} \cline{2-2} \cline{4-4} \cline{5-5} \cline{6-6} \cline{7-7} \cline{8-8} \cline{9-9} \cline{10-10}
\multicolumn{1}{c}{} & \multicolumn{1}{c|}{} & \multicolumn{1}{c}{{\footnotesize 15}} & {\footnotesize 14} & {\footnotesize 13} & {\footnotesize 12} & {\footnotesize 11} & {\footnotesize 10} & \multicolumn{1}{c|}{{\footnotesize 9}} & \multicolumn{1}{c|}{{\footnotesize $l\,\diagdown n$}}\tabularnewline
\cline{3-3} \cline{4-4} \cline{5-5} \cline{6-6} \cline{7-7} \cline{8-8} \cline{9-9} \cline{10-10}
\end{tabular}
\label{T2}
\end{center}
\end{table}

\begin{table}
\begin{center}
\caption{Critical parameters $\beta=2mgr_0\hbar^{-2}$ of the Yukawa potential $V(r)=-g~\exp(-r/r_0)/r$.}
\begin{tabular}{|c|ccccccccc|}
\hline
{\footnotesize $n\diagdown l$} & {\footnotesize 0} & {\footnotesize 1} & {\footnotesize 2} & {\footnotesize 3} & {\footnotesize 4} & {\footnotesize 5} & {\footnotesize 6} & {\footnotesize 7} & {\footnotesize 8}\tabularnewline
\hline
{\footnotesize 1} & {\footnotesize 1.6798078} & {\footnotesize 9.0819590} & {\footnotesize 21.894984} & {\footnotesize 40.135552} & {\footnotesize 63.808976} & {\footnotesize 92.917164} & {\footnotesize 127.46092} & {\footnotesize 167.44064} & {\footnotesize 212.85654}\tabularnewline
{\footnotesize 2} & {\footnotesize 6.4472603} & {\footnotesize 17.744576} & {\footnotesize 34.420414} & {\footnotesize 56.514114} & {\footnotesize 84.036777} & {\footnotesize 116.99234} & {\footnotesize 155.38248} & {\footnotesize 199.20797} & {\footnotesize 248.46926}\tabularnewline
{\footnotesize 3} & {\footnotesize 14.342028} & {\footnotesize 29.461426} & {\footnotesize 49.969576} & {\footnotesize 75.899394} & {\footnotesize 107.26037} & {\footnotesize 144.05569} & {\footnotesize 186.28656} & {\footnotesize 233.95349} & {\footnotesize 287.05673}\tabularnewline
{\footnotesize 4} & {\footnotesize 25.371660} & {\footnotesize 44.261254} & {\footnotesize 68.571467 } & {\footnotesize 98.317925} & {\footnotesize 133.50366} & {\footnotesize 174.12875} & {\footnotesize 220.19272} & {\footnotesize 271.69505} & {\footnotesize 328.63535}\tabularnewline
{\footnotesize 5} & {\footnotesize 39.538842} & {\footnotesize 62.160193} & {\footnotesize 90.245270} & {\footnotesize 123.78892} & {\footnotesize 162.78498} & {\footnotesize 207.22876} & {\footnotesize 257.11705} & {\footnotesize 312.44769} & {\footnotesize 373.21919}\tabularnewline
{\footnotesize 6} & {\footnotesize 56.84486} & {\footnotesize 83.168247} & {\footnotesize 115.00434} & {\footnotesize 152.32675} & {\footnotesize 195.11871} & {\footnotesize 243.36968} & {\footnotesize 297.07295} & {\footnotesize 356.22417} & {\footnotesize 420.82039}\tabularnewline
{\footnotesize 7} & {\footnotesize 77.290455} & {\footnotesize 107.29208} & {\footnotesize 142.85836 } & {\footnotesize 183.94242} & {\footnotesize 230.51632} & {\footnotesize 282.56299} & {\footnotesize 340.07168} & {\footnotesize 403.03541} & {\footnotesize 471.44949}\tabularnewline
{\footnotesize 8} & {\footnotesize 100.87607} & {\footnotesize 134.53636} & {\footnotesize 173.81459} & {\footnotesize 218.64457} & {\footnotesize 268.98711} & {\footnotesize 324.81824} & {\footnotesize 386.12280} & {\footnotesize 452.89086} & {\footnotesize 525.11571}\tabularnewline
\cline{10-10}
{\footnotesize 9} & {\footnotesize 127.60202} & {\footnotesize 164.90453 } & {\footnotesize 207.87862} & {\footnotesize 256.44012} & {\footnotesize 310.53874 } & {\footnotesize 370.14347} & {\footnotesize 435.23449} & \multicolumn{1}{c|}{{\footnotesize 505.79870 }} & \multicolumn{1}{c|}{}\tabularnewline
\cline{9-9}
{\footnotesize 10} & {\footnotesize 157.46853} & {\footnotesize 198.39917} & {\footnotesize 245.05485} & {\footnotesize 297.33466} & {\footnotesize 355.17758 } & {\footnotesize 418.54550} & \multicolumn{1}{c|}{{\footnotesize 487.41381}} & \multicolumn{1}{c|}{{\footnotesize 545.31066}} & \multicolumn{1}{c|}{{\footnotesize 7}}\tabularnewline
\cline{8-8}
{\footnotesize 11} & {\footnotesize 190.47575} & {\footnotesize 235.02231} & {\footnotesize 285.34681} & {\footnotesize 341.33284} & {\footnotesize 402.90900} & \multicolumn{1}{c|}{{\footnotesize 470.03018 }} & \multicolumn{1}{c}{{\footnotesize 566.34038}} & \multicolumn{1}{c|}{{\footnotesize 490.8596}} & \multicolumn{1}{c|}{{\footnotesize 6}}\tabularnewline
\cline{7-7}
{\footnotesize 12} & {\footnotesize 226.62381} & {\footnotesize 274.77554} & {\footnotesize 328.75740} & {\footnotesize 388.43851} & \multicolumn{1}{c|}{{\footnotesize 453.73757 }} & \multicolumn{1}{c}{{\footnotesize 588.17003}} & {\footnotesize 511.08099} & \multicolumn{1}{c|}{{\footnotesize 439.43054}} & \multicolumn{1}{c|}{{\footnotesize 5}}\tabularnewline
\cline{6-6}
{\footnotesize 13} & {\footnotesize 265.91281} & {\footnotesize 317.66016} & {\footnotesize 375.28899} & \multicolumn{1}{c|}{{\footnotesize 438.65494}} & \multicolumn{1}{c}{{\footnotesize 610.77114}} & {\footnotesize 532.08136} & {\footnotesize 458.82871} & \multicolumn{1}{c|}{{\footnotesize 391.01331}} & \multicolumn{1}{c|}{{\footnotesize 4}}\tabularnewline
\cline{5-5}
{\footnotesize 14} & {\footnotesize 308.34282} & {\footnotesize 363.67724} & \multicolumn{1}{c|}{{\footnotesize 424.94360}} & \multicolumn{1}{c}{{\footnotesize 634.12021}} & {\footnotesize 553.83444} & {\footnotesize 478.98522} & {\footnotesize 409.57254} & \multicolumn{1}{c|}{{\footnotesize 345.59639}} & \multicolumn{1}{c|}{{\footnotesize 3}}\tabularnewline
\cline{4-4}
{\footnotesize 15} & {\footnotesize 353.91391} & \multicolumn{1}{c|}{{\footnotesize 412.82768}} & \multicolumn{1}{c}{{\footnotesize 658.19765}} & {\footnotesize 576.31864} & {\footnotesize 499.87607} & {\footnotesize 428.86991} & {\footnotesize 363.30010} & \multicolumn{1}{c|}{{\footnotesize 303.16658}} & \multicolumn{1}{c|}{{\footnotesize 2}}\tabularnewline
\cline{3-3}
{\footnotesize 16} & \multicolumn{1}{c|}{{\footnotesize 402.62614}} & \multicolumn{1}{c}{{\footnotesize 682.98698}} & {\footnotesize 599.51605} & {\footnotesize 521.48163} & {\footnotesize 448.88371} & {\footnotesize 381.72227} & {\footnotesize 319.99729} & \multicolumn{1}{c|}{{\footnotesize 263.70873}} & \multicolumn{1}{c|}{{\footnotesize 1}}\tabularnewline
\cline{3-3}
\cline{1-1} \cline{2-2} \cline{4-4} \cline{5-5} \cline{6-6} \cline{7-7} \cline{8-8} \cline{9-9} \cline{10-10}
\multicolumn{1}{c}{} & \multicolumn{1}{c|}{} & \multicolumn{1}{c}{{\footnotesize 15}} & {\footnotesize 14} & {\footnotesize 13} & {\footnotesize 12} & {\footnotesize 11} & {\footnotesize 10} & \multicolumn{1}{c|}{{\footnotesize 9}} & \multicolumn{1}{c|}{{\footnotesize $l\,\diagdown n$}}\tabularnewline
\cline{3-3} \cline{4-4} \cline{5-5} \cline{6-6} \cline{7-7} \cline{8-8} \cline{9-9} \cline{10-10}
\end{tabular}
\label{T3}
\end{center}
\end{table}

\begin{table}
\begin{center}
\caption{Critical parameters $\beta=2mgr_0^2\hbar^{-2}$ of the Gaussian potential $V(r)=-g~\exp(-r^2/r_0^2)$.}
\begin{tabular}{|c|ccccccccc|}
\hline
{\footnotesize $n\diagdown l$} & {\footnotesize 0} & {\footnotesize 1} & {\footnotesize 2} & {\footnotesize 3} & {\footnotesize 4} & {\footnotesize 5} & {\footnotesize 6} & {\footnotesize 7} & {\footnotesize 8}\tabularnewline
\hline
{\footnotesize 1} & {\footnotesize 2.6840047} & {\footnotesize 12.099309} & {\footnotesize 26.901078} & {\footnotesize 47.107862} & {\footnotesize 72.733004} & {\footnotesize 103.78395} & {\footnotesize 140.26479} & {\footnotesize 182.17789} & {\footnotesize 229.52463}\tabularnewline
{\footnotesize 2} & {\footnotesize 17.795700} & {\footnotesize 35.089777} & {\footnotesize 57.675772} & {\footnotesize 85.627393} & {\footnotesize 118.97711} & {\footnotesize 157.74107} & {\footnotesize 201.92788} & {\footnotesize 251.54239} & {\footnotesize 306.58749}\tabularnewline
{\footnotesize 3} & {\footnotesize 45.573480} & {\footnotesize 70.482856} & {\footnotesize 100.71505} & {\footnotesize 136.32500 } & {\footnotesize 177.33962} & {\footnotesize 223.77276} & {\footnotesize 275.63189} & {\footnotesize 332.92115} & {\footnotesize 395.64293}\tabularnewline
{\footnotesize 4} & {\footnotesize 85.963400} & {\footnotesize 118.35165} & {\footnotesize 156.12749} & {\footnotesize 199.31846} & {\footnotesize 247.93811} & {\footnotesize 301.99288} & {\footnotesize 361.48565} & {\footnotesize 426.41761} & {\footnotesize 496.78909}\tabularnewline
{\footnotesize 5} & {\footnotesize 138.94811} & {\footnotesize 178.72771} & {\footnotesize 223.97020} & {\footnotesize 274.67746} & {\footnotesize 330.84800} & {\footnotesize 392.47880} & {\footnotesize 459.56651} & {\footnotesize 532.10794} & {\footnotesize 610.10035}\tabularnewline
{\footnotesize 6} & {\footnotesize 204.51926} & {\footnotesize 251.62783} & {\footnotesize 304.27712} & {\footnotesize 362.44675} & {\footnotesize 426.12056} & {\footnotesize 495.28556} & {\footnotesize 569.93140} & {\footnotesize 650.04986} & {\footnotesize 735.63442}\tabularnewline
{\footnotesize 7} & {\footnotesize 282.67201} & {\footnotesize 337.06209} & {\footnotesize 397.07024} & {\footnotesize 462.65690} & {\footnotesize 533.79230} & {\footnotesize 610.45368} & {\footnotesize 692.62349} & {\footnotesize 780.28811} & {\footnotesize 873.43695 }\tabularnewline
{\footnotesize 8} & {\footnotesize 373.40327} & {\footnotesize 435.03714} & {\footnotesize 502.36469} & {\footnotesize 575.32978} & {\footnotesize 653.8902} & {\footnotesize 738.01395} & {\footnotesize 827.67628} & {\footnotesize 922.85815} & {\footnotesize 1023.5447}\tabularnewline
\cline{10-10}
{\footnotesize 9} & {\footnotesize 476.71088} & {\footnotesize 545.55758 } & {\footnotesize 620.17137} & {\footnotesize 700.48163} & {\footnotesize 786.43489} & {\footnotesize 877.99030} & {\footnotesize 975.11633} & \multicolumn{1}{c|}{{\footnotesize 1077.7885}} & \multicolumn{1}{c|}{}\tabularnewline
\cline{9-9}
{\footnotesize 10} & {\footnotesize 592.59331} & {\footnotesize 668.62678} & {\footnotesize 750.49845} & {\footnotesize 838.12487} & {\footnotesize 931.44233} & {\footnotesize 1030.4018} & \multicolumn{1}{c|}{{\footnotesize 1134.9651}} & \multicolumn{1}{c|}{{\footnotesize 972.06166}} & \multicolumn{1}{c|}{{\footnotesize 7}}\tabularnewline
\cline{8-8}
{\footnotesize 11} & {\footnotesize 721.04936} & {\footnotesize 804.24729} & {\footnotesize 893.35221} & {\footnotesize 988.26926} & {\footnotesize 1088.9253} & \multicolumn{1}{c|}{{\footnotesize 1195.2637}} & \multicolumn{1}{c}{{\footnotesize 923.18228}} & \multicolumn{1}{c|}{{\footnotesize 826.67993}} & \multicolumn{1}{c|}{{\footnotesize 6}}\tabularnewline
\cline{7-7}
{\footnotesize 12} & {\footnotesize 862.07814} & {\footnotesize 952.42108} & {\footnotesize 1048.7376} & {\footnotesize 1150.9226} & \multicolumn{1}{c|}{{\footnotesize 1258.8943}} & \multicolumn{1}{c}{{\footnotesize 876.76204}} & {\footnotesize 782.42916} & \multicolumn{1}{c|}{{\footnotesize 693.54140}} & \multicolumn{1}{c|}{{\footnotesize 5}}\tabularnewline
\cline{6-6}
{\footnotesize 13} & {\footnotesize 1015.6789} & {\footnotesize 1113.1497} & {\footnotesize 1216.6587} & \multicolumn{1}{c|}{{\footnotesize 1326.0912}} & \multicolumn{1}{c}{{\footnotesize 832.66745}} & {\footnotesize 740.53949} & {\footnotesize 653.85026} & \multicolumn{1}{c|}{{\footnotesize 572.60004}} & \multicolumn{1}{c|}{{\footnotesize 4}}\tabularnewline
\cline{5-5}
{\footnotesize 14} & {\footnotesize 1181.8511} & {\footnotesize 1286.4345} & \multicolumn{1}{c|}{{\footnotesize 1397.1187}} & \multicolumn{1}{c}{{\footnotesize 790.77458}} & {\footnotesize 700.87689} & {\footnotesize 616.41518} & {\footnotesize 537.38923} & \multicolumn{1}{c|}{{\footnotesize 463.79866}} & \multicolumn{1}{c|}{{\footnotesize 3}}\tabularnewline
\cline{4-4}
{\footnotesize 15} & {\footnotesize 1360.5941} & \multicolumn{1}{c|}{{\footnotesize 1472.2765}} & \multicolumn{1}{c}{{\footnotesize 750.96802}} & {\footnotesize 663.31706} & {\footnotesize 581.10160} & {\footnotesize 504.32139} & {\footnotesize 432.97603} & \multicolumn{1}{c|}{{\footnotesize 367.06499}} & \multicolumn{1}{c|}{{\footnotesize 2}}\tabularnewline
\cline{3-3}
{\footnotesize 16} & \multicolumn{1}{c|}{{\footnotesize 1551.9077}} & \multicolumn{1}{c}{{\footnotesize 713.13988}} & {\footnotesize 627.74431} & {\footnotesize 547.78488} & {\footnotesize 473.26149} & {\footnotesize 404.17400} & {\footnotesize 340.52222} & \multicolumn{1}{c|}{{\footnotesize 282.30589}} & \multicolumn{1}{c|}{{\footnotesize 1}}\tabularnewline
\cline{3-3}
\cline{1-1} \cline{2-2} \cline{4-4} \cline{5-5} \cline{6-6} \cline{7-7} \cline{8-8} \cline{9-9} \cline{10-10}
\multicolumn{1}{c}{} & \multicolumn{1}{c|}{} & \multicolumn{1}{c}{{\footnotesize 15}} & {\footnotesize 14} & {\footnotesize 13} & {\footnotesize 12} & {\footnotesize 11} & {\footnotesize 10} & \multicolumn{1}{c|}{{\footnotesize 9}} & \multicolumn{1}{c|}{{\footnotesize $l\,\diagdown n$}}\tabularnewline
\cline{3-3} \cline{4-4} \cline{5-5} \cline{6-6} \cline{7-7} \cline{8-8} \cline{9-9} \cline{10-10}
\end{tabular}
\label{T4}
\end{center}
\end{table}

\begin{table}
\begin{center}
\caption{Critical parameters $\beta=2mgr_0^2\hbar^{-2}$ of the Woods-Saxon potential~\\
 $V(r)=-g/[1+\exp(r / r_0 )]$.}
\begin{tabular}{|c|ccccccccc|}
\hline
{\footnotesize $n\diagdown l$} & {\footnotesize 0} & {\footnotesize 1} & {\footnotesize 2} & {\footnotesize 3} & {\footnotesize 4} & {\footnotesize 5} & {\footnotesize 6} & {\footnotesize 7} & {\footnotesize 8}\tabularnewline
\hline
{\footnotesize 1} & {\footnotesize 1.7205730} & {\footnotesize 8.2135317} & {\footnotesize 18.813940} & {\footnotesize 33.542490} & {\footnotesize 52.406813} & {\footnotesize 75.410024} & {\footnotesize 102.553558} & {\footnotesize 133.83814} & {\footnotesize 169.26419}\tabularnewline
{\footnotesize 2} & {\footnotesize 9.6742198} & {\footnotesize 20.755782} & {\footnotesize 35.931282} & {\footnotesize 55.231021} & {\footnotesize 78.665326} & {\footnotesize 106.23821} & {\footnotesize 137.9514072} & {\footnotesize 173.80576} & {\footnotesize 213.80169}\tabularnewline
{\footnotesize 3} & {\footnotesize 23.969284} & {\footnotesize 39.431300} & {\footnotesize 59.073131} & {\footnotesize 82.876856} & {\footnotesize 110.83541} & {\footnotesize 142.94481 } & {\footnotesize 179.2025315} & {\footnotesize 219.60692} & {\footnotesize 264.15685 }\tabularnewline
{\footnotesize 4} & {\footnotesize 44.615717} & {\footnotesize 64.332441} & {\footnotesize 88.346797} & {\footnotesize 116.58590} & {\footnotesize 149.01715} & {\footnotesize 185.62315} & {\footnotesize 226.3936688} & {\footnotesize 271.32232} & {\footnotesize 320.40492}\tabularnewline
{\footnotesize 5} & {\footnotesize 71.614405} & {\footnotesize 95.502170} & {\footnotesize 123.81377} & {\footnotesize 156.42640} & {\footnotesize 193.28014} & {\footnotesize 234.34169 } & {\footnotesize 279.5910514} & {\footnotesize 329.01551} & {\footnotesize 382.60663}\tabularnewline
{\footnotesize 6} & {\footnotesize 104.96555} & {\footnotesize 132.96409} & {\footnotesize 165.51246} & {\footnotesize 202.44471} & {\footnotesize 243.67452} & {\footnotesize 289.15197} & {\footnotesize 338.8462258} & {\footnotesize 392.73727} & {\footnotesize 450.81159}\tabularnewline
{\footnotesize 7} & {\footnotesize 144.66921} & {\footnotesize 176.73264} & {\footnotesize 213.46842} & {\footnotesize 254.67366} & {\footnotesize 300.23751} & {\footnotesize 350.09364} & {\footnotesize 404.2000006} & {\footnotesize 462.52874 } & {\footnotesize 525.06071}\tabularnewline
{\footnotesize 8} & {\footnotesize 190.72542} & {\footnotesize 226.81735 } & {\footnotesize 267.69958} & {\footnotesize 313.13737} & {\footnotesize 362.99747} & {\footnotesize 417.19781} & {\footnotesize 475.6851818} & {\footnotesize 538.42365} & {\footnotesize 605.38815 }\tabularnewline
\cline{10-10}
{\footnotesize 9} & {\footnotesize 243.13418} & {\footnotesize 283.22488} & {\footnotesize 328.218988} & {\footnotesize 377.85407} & {\footnotesize 431.97649} & {\footnotesize 490.48934} & {\footnotesize 553.3285074} & \multicolumn{1}{c|}{{\footnotesize 620.44998}} & \multicolumn{1}{c|}{}\tabularnewline
\cline{9-9}
{\footnotesize 10} & {\footnotesize 301.89551} & {\footnotesize 345.96008} & {\footnotesize 395.03646 } & {\footnotesize 448.83786} & {\footnotesize 507.19210} & {\footnotesize 569.98836} & \multicolumn{1}{c|}{{\footnotesize 637.1520405}} & \multicolumn{1}{c|}{{\footnotesize 591.78238}} & \multicolumn{1}{c|}{{\footnotesize 7}}\tabularnewline
\cline{8-8}
{\footnotesize 11} & {\footnotesize 367.00940} & {\footnotesize 415.02662} & {\footnotesize 468.15961} & {\footnotesize 526.09990} & {\footnotesize 588.65845} & \multicolumn{1}{c|}{{\footnotesize 655.71141}} & \multicolumn{1}{c}{{\footnotesize 579.47499}} & \multicolumn{1}{c|}{{\footnotesize 513.05975}} & \multicolumn{1}{c|}{{\footnotesize 6}}\tabularnewline
\cline{7-7}
{\footnotesize 12} & {\footnotesize 438.47587} & {\footnotesize 490.42734} & {\footnotesize 547.59441} & {\footnotesize 609.64918} & \multicolumn{1}{c|}{{\footnotesize 676.38713}} & \multicolumn{1}{c}{{\footnotesize 568.32987}} & {\footnotesize 502.26737} & \multicolumn{1}{c|}{{\footnotesize 440.35862}} & \multicolumn{1}{c|}{{\footnotesize 5}}\tabularnewline
\cline{6-6}
{\footnotesize 13} & {\footnotesize 516.29490 } & {\footnotesize 572.16450} & {\footnotesize 633.34572} & \multicolumn{1}{c|}{{\footnotesize 699.49305}} & \multicolumn{1}{c}{{\footnotesize 558.22885}} & {\footnotesize 492.55183} & {\footnotesize 431.02140} & \multicolumn{1}{c|}{{\footnotesize 373.63861}} & \multicolumn{1}{c|}{{\footnotesize 4}}\tabularnewline
\cline{5-5}
{\footnotesize 14} & {\footnotesize 600.46650} & {\footnotesize 660.23992} & \multicolumn{1}{c|}{{\footnotesize 725.41749}} & \multicolumn{1}{c}{{\footnotesize 549.06873}} & {\footnotesize 483.79929} & {\footnotesize 422.67311} & {\footnotesize 365.69043} & \multicolumn{1}{c|}{{\footnotesize 312.85154}} & \multicolumn{1}{c|}{{\footnotesize 3}}\tabularnewline
\cline{4-4}
{\footnotesize 15} & {\footnotesize 690.99068} & \multicolumn{1}{c|}{{\footnotesize 754.65510}} & \multicolumn{1}{c}{{\footnotesize 540.75901}} & {\footnotesize 475.91083} & {\footnotesize 415.20481} & {\footnotesize 358.64093} & {\footnotesize 306.21917} & \multicolumn{1}{c|}{{\footnotesize 257.93944}} & \multicolumn{1}{c|}{{\footnotesize 2}}\tabularnewline
\cline{3-3}
{\footnotesize 16} & \multicolumn{1}{c|}{{\footnotesize 787.86743}} & \multicolumn{1}{c}{{\footnotesize 533.21997}} & {\footnotesize 468.80008} & {\footnotesize 408.52232} & {\footnotesize 352.38667} & {\footnotesize 300.39309} & {\footnotesize 252.54154} & \multicolumn{1}{c|}{{\footnotesize 208.83194 }} & \multicolumn{1}{c|}{{\footnotesize 1}}\tabularnewline
\cline{3-3}
\cline{1-1} \cline{2-2} \cline{4-4} \cline{5-5} \cline{6-6} \cline{7-7} \cline{8-8} \cline{9-9} \cline{10-10}
\multicolumn{1}{c}{} & \multicolumn{1}{c|}{} & \multicolumn{1}{c}{{\footnotesize 15}} & {\footnotesize 14} & {\footnotesize 13} & {\footnotesize 12} & {\footnotesize 11} & {\footnotesize 10} & \multicolumn{1}{c|}{{\footnotesize 9}} & \multicolumn{1}{c|}{{\footnotesize $l\,\diagdown n$}}\tabularnewline
\cline{3-3} \cline{4-4} \cline{5-5} \cline{6-6} \cline{7-7} \cline{8-8} \cline{9-9} \cline{10-10}
\end{tabular}
\label{T5}
\end{center}
\end{table}

\begin{table}
\begin{center}
\caption{Critical parameters $\beta=2mgr_0^2\hbar^{-2}$ of the Woods-Saxon potential~\\
 $V(r)=-g/[1+0.001\exp(r / r_0 )]$.}
\begin{tabular}{|c|ccccccccc|}
\hline
{\footnotesize $n\diagdown l$} & {\footnotesize 0} & {\footnotesize 1} & {\footnotesize 2} & {\footnotesize 3} & {\footnotesize 4} & {\footnotesize 5} & {\footnotesize 6} & {\footnotesize 7} & {\footnotesize 8}\tabularnewline
\hline
{\footnotesize 1} & {\footnotesize .050143914} & {\footnotesize .21236959} & {\footnotesize .45411555} & {\footnotesize .77426459} & {\footnotesize 1.1724439} & {\footnotesize 1.6484948} & {\footnotesize 2.2023382} & {\footnotesize 2.8339302} & {\footnotesize 3.5432443}\tabularnewline
{\footnotesize 2} & {\footnotesize .40525072} & {\footnotesize .74663148} & {\footnotesize 1.1653050} & {\footnotesize 1.6617334} & {\footnotesize 2.2359950} & {\footnotesize 2.8880871} & {\footnotesize 3.6179882} & {\footnotesize 4.4256753} & {\footnotesize 5.3111276}\tabularnewline
{\footnotesize 3} & {\footnotesize 1.0529125} & {\footnotesize 1.5665339} & {\footnotesize 2.1583382} & {\footnotesize 2.8284741} & {\footnotesize 3.5768855} & {\footnotesize 4.4034793} & {\footnotesize 5.3081667} & {\footnotesize 6.2908725 } & {\footnotesize 7.3515353}\tabularnewline
{\footnotesize 4} & {\footnotesize 1.9863174} & {\footnotesize 2.6709161} & {\footnotesize 3.4345909} & {\footnotesize 4.2772382} & {\footnotesize 5.1986738} & {\footnotesize 6.1987162} & {\footnotesize 7.2772089} & {\footnotesize 8.4340224} & {\footnotesize 9.6690512}\tabularnewline
{\footnotesize 5} & {\footnotesize 3.2060640} & {\footnotesize 4.0606118} & {\footnotesize 4.9951932} & {\footnotesize 6.0094520} & {\footnotesize 7.1030599} & {\footnotesize 8.2757388} & {\footnotesize 9.5272608} & {\footnotesize 10.857441} & {\footnotesize 12.266128}\tabularnewline
{\footnotesize 6} & {\footnotesize 4.7125157} & {\footnotesize 5.7361307} & {\footnotesize 6.8408115} & {\footnotesize 8.0259370} & {\footnotesize 9.2910171} & {\footnotesize 10.635665} & {\footnotesize 12.059576} & {\footnotesize 13.562505} & {\footnotesize 15.144255}\tabularnewline
{\footnotesize 7} & {\footnotesize 6.5057952} & {\footnotesize 7.6977523} & {\footnotesize 8.9718549} & {\footnotesize 10.327215} & {\footnotesize 11.763171} & {\footnotesize 13.279221} & {\footnotesize 14.874975 } & {\footnotesize 16.550125} & {\footnotesize 18.304425}\tabularnewline
{\footnotesize 8} & {\footnotesize 8.5859457} & {\footnotesize 9.9456459} & {\footnotesize 11.388595} & {\footnotesize 12.913644 } & {\footnotesize 14.519959} & {\footnotesize 16.206914} & {\footnotesize 17.974033 } & {\footnotesize 19.820942} & {\footnotesize 21.747345}\tabularnewline
\cline{10-10}
{\footnotesize 9} & {\footnotesize 10.952985} & {\footnotesize 12.479924} & {\footnotesize 14.091224} & {\footnotesize 15.785486} & {\footnotesize 17.561701} & {\footnotesize 19.419123} & {\footnotesize 21.357183} & \multicolumn{1}{c|}{{\footnotesize 23.375440 }} & \multicolumn{1}{c|}{}\tabularnewline
\cline{9-9}
{\footnotesize 10} & {\footnotesize 13.606920} & {\footnotesize 15.300668} & {\footnotesize 17.079885} & {\footnotesize 18.942937} & {\footnotesize 20.888645} & {\footnotesize 22.916139} & \multicolumn{1}{c|}{{\footnotesize 25.024758}} & \multicolumn{1}{c|}{{\footnotesize 20.137676}} & \multicolumn{1}{c|}{{\footnotesize 7}}\tabularnewline
\cline{8-8}
{\footnotesize 11} & {\footnotesize 16.547758} & {\footnotesize 18.407936} & {\footnotesize 20.354690} & {\footnotesize 22.386154} & {\footnotesize 24.500987} & \multicolumn{1}{c|}{{\footnotesize 26.698194}} & \multicolumn{1}{c}{{\footnotesize 18.543603}} & \multicolumn{1}{c|}{{\footnotesize 16.804665}} & \multicolumn{1}{c|}{{\footnotesize 6}}\tabularnewline
\cline{7-7}
{\footnotesize 12} & {\footnotesize 19.775499} & {\footnotesize 21.801777} & {\footnotesize 23.915725} & {\footnotesize 26.115262} & \multicolumn{1}{c|}{{\footnotesize 28.398884}} & \multicolumn{1}{c}{{\footnotesize 16.962111}} & {\footnotesize 15.318555} & \multicolumn{1}{c|}{{\footnotesize 13.753200}} & \multicolumn{1}{c|}{{\footnotesize 5}}\tabularnewline
\cline{6-6}
{\footnotesize 13} & {\footnotesize 23.290146} & {\footnotesize 25.482228} & {\footnotesize 27.763062} & \multicolumn{1}{c|}{{\footnotesize 30.130362}} & \multicolumn{1}{c}{{\footnotesize 15.389823}} & {\footnotesize 13.842647} & {\footnotesize 12.373427} & \multicolumn{1}{c|}{{\footnotesize 10.982209}} & \multicolumn{1}{c|}{{\footnotesize 4}}\tabularnewline
\cline{5-5}
{\footnotesize 14} & {\footnotesize 27.091699} & {\footnotesize 29.449318} & \multicolumn{1}{c|}{{\footnotesize 31.896761 }} & \multicolumn{1}{c}{{\footnotesize 13.822777}} & {\footnotesize 12.372904} & {\footnotesize 11.000819} & {\footnotesize 9.7065447} & \multicolumn{1}{c|}{{\footnotesize 8.4901060}} & \multicolumn{1}{c|}{{\footnotesize 3}}\tabularnewline
\cline{4-4}
{\footnotesize 15} & {\footnotesize 31.180159} & \multicolumn{1}{c|}{{\footnotesize 33.703074}} & \multicolumn{1}{c}{{\footnotesize 12.256127}} & {\footnotesize 10.904356} & {\footnotesize 9.6302842} & {\footnotesize 8.4339165} & {\footnotesize 7.3152612} & \multicolumn{1}{c|}{{\footnotesize 6.2743276}} & \multicolumn{1}{c|}{{\footnotesize 2}}\tabularnewline
\cline{3-3}
{\footnotesize 16} & \multicolumn{1}{c|}{{\footnotesize 35.555527}} & \multicolumn{1}{c}{{\footnotesize 10.683731}} & {\footnotesize 9.4306340} & {\footnotesize 8.2552072} & {\footnotesize 7.1574531} & {\footnotesize 6.1373746} & {\footnotesize 5.1949761} & \multicolumn{1}{c|}{{\footnotesize 4.3302634}} & \multicolumn{1}{c|}{{\footnotesize 1}}\tabularnewline
\cline{3-3}
\cline{1-1} \cline{2-2} \cline{4-4} \cline{5-5} \cline{6-6} \cline{7-7} \cline{8-8} \cline{9-9} \cline{10-10}
\multicolumn{1}{c}{} & \multicolumn{1}{c|}{} & \multicolumn{1}{c}{{\footnotesize 15}} & {\footnotesize 14} & {\footnotesize 13} & {\footnotesize 12} & {\footnotesize 11} & {\footnotesize 10} & \multicolumn{1}{c|}{{\footnotesize 9}} & \multicolumn{1}{c|}{{\footnotesize $l\,\diagdown n$}}\tabularnewline
\cline{3-3} \cline{4-4} \cline{5-5} \cline{6-6} \cline{7-7} \cline{8-8} \cline{9-9} \cline{10-10}
\end{tabular}
\label{T6}
\end{center}
\end{table}

\end{document}